\documentclass[aps,prb,reprint,amssymb,superscriptaddress]{revtex4-1}

\usepackage{amsmath,amssymb,color}
\usepackage{bm}
\usepackage{hyperref}
 \usepackage{graphicx}

\newcommand{\kv}{{\bm{k}}}
\newcommand{\qv}{{\bm{q}}}

\begin{document}

\title{Accumulation of  magnetoelastic bosons in yttrium iron garnet: 
kinetic theory and wave vector resolved Brillouin light scattering}

\date{\today}

\author{Viktor Hahn}
\affiliation{Institut f\"{u}r Theoretische Physik, Universit\"{a}t Frankfurt, Max-von-Laue Stra{\ss}e 1, 60438 Frankfurt, Germany}
\email{hahn@itp.uni-frankfurt.de}

\author{Pascal Frey}
\affiliation{Fachbereich Physik and Landesforschungszentrum OPTIMAS, Technische Universit\"{a}t Kaiserslautern, 67663 Kaiserslautern, Germany}
\author{Alexander A. Serga}
\affiliation{Fachbereich Physik and Landesforschungszentrum OPTIMAS, Technische Universit\"{a}t Kaiserslautern, 67663 Kaiserslautern, Germany}
\author{Vitaliy I. Vasyuchka}
\affiliation{Fachbereich Physik and Landesforschungszentrum OPTIMAS, Technische Universit\"{a}t Kaiserslautern, 67663 Kaiserslautern, Germany}
\author{Burkard Hillebrands}
\affiliation{Fachbereich Physik and Landesforschungszentrum OPTIMAS, Technische Universit\"{a}t Kaiserslautern, 67663 Kaiserslautern, Germany}

\author{Peter Kopietz}
\affiliation{Institut f\"{u}r Theoretische Physik, Universit\"{a}t Frankfurt, Max-von-Laue Stra{\ss}e 1, 60438 Frankfurt, Germany}
\affiliation{Department of Physics and Astronomy,
University of California, Irvine, California 92697, USA}
\author{Andreas R\"{u}ckriegel}
\affiliation{Institut f\"{u}r Theoretische Physik, Universit\"{a}t Frankfurt, Max-von-Laue Stra{\ss}e 1, 60438 Frankfurt, Germany}

\begin{abstract} 
We derive and solve
quantum kinetic equations describing the accumulation of magnetoelastic bosons
in an overpopulated magnon gas realized in a thin film of the 
magnetic insulator yttrium iron garnet.
We show that in the presence of a magnon condensate,
there is a non-equilibrium steady state in which incoherent magnetoelastic bosons
accumulate in a narrow region in momentum space for
energies slightly below the bottom of the magnon spectrum.
The results of our calculations agree quite well with  
Brillouin light scattering measurements of the  stationary non-equilibrium state
of magnons and magnetoelastic bosons in  
yttrium iron garnet. 
\end{abstract}

\maketitle

\section{Introduction}

The theoretical investigation of 
magnon-phonon interactions in magnetic insulators was initiated by
Abrahams and Kittel~\cite{Abrahams52} in 1952.
Over the years   the interest in this topic has waned and often
the effect of the phonons on magnons
has been  taken into account only
implicitly by assuming that phonons merely serve as a thermal bath and an energy sink for the magnons.
Recently magnon-phonon interactions have attracted renewed interest  
in the field of 
spintronics \cite{Bozhko20} where one can now study phenomena
which are dominated by the magnetoelastic coupling.\cite{Kamra14,Rueckriegel14,Ogawa15,Kikkawa16,Takahashi16,Baryakhtar17,Ramos19,Rueckriegel20}
Of particular interest are magnetoelastic bosons which  emerge 
because of the hybridization of magnons with  phonons and as such combine properties of both. For example, in a recent series of experiments,~\cite{Bozhko17,Frey21}
the spontaneous accumulation of magnetoelastic bosons 
during the thermalization of an overpopulated magnon gas in the magnetic insulator
yttrium iron garnet (YIG) was observed
by Brillouin light scattering spectroscopy.
While a phenomenological explanation of this observation
in terms of a bottleneck accumulation effect was already provided by the authors of Ref.~[\onlinecite{Bozhko17}],
important questions about the nature of the accumulation remain open.
For example,
it is not clear whether the accumulation in the magnetoelastic mode is coherent.
Moreover, in addition to the magnetoelastic accumulation, in the experiment 
a magnon condensate at the bottom of the 
magnon spectrum was also observed.
As the magnon condensate and the magnetoelastic boson  
are energetically nearly degenerate,
this raises the question of the importance of interactions between these different
types of modes.

The theory of magnons, phonons, and hybrid magnetoelastic bosons in YIG films 
is already well-developed, 
see for example Refs.~[\onlinecite{Kalinkos86,Kreisel09,Rueckriegel14}]. 
In the present work,
we go beyond this established theory by developing a kinetic theory 
for the coupled magnon-phonon system 
which allows us to gain a complete  microscopic understanding 
of the physical processes leading to the accumulation of magnetoelastic bosons in YIG~\cite{Bozhko17,Frey21}.
To this end, we derive quantum kinetic equations for the incoherent distribution functions and condensate amplitudes
of the magnetoelastic bosons.
We then solve the  kinetic equations numerically
to obtain a non-equilibrium steady state
that displays the magnetoelastic accumulation. In the experimental section of 
this work, we present  new wave vector resolved Brillouin light scattering 
results for the magnetoelastic accumulation in YIG
which are in good agreement with our theoretical predictions.

The rest of this work  is organized as follows:
In Sec.~\ref{sec:bosons},
we briefly review the theory of magnons and phonons in thin YIG films; in particular, we
discuss the magnetoelastic modes and the relevant interaction vertices.
The quantum kinetic equations describing the dynamics of the coupled magnon-phonon system
are derived and self-consistently solved
in Sec.~\ref{sec:Accumulation}; we also
compare our theoretical results 
with new Brillouin light scattering measurements.
In the concluding  Sec.~\ref{sec:conclusion} we  briefly summarize our results.
Finally, in three appendices we present technical details
of the derivation of the magnon-phonon Hamiltonian in YIG and of 
the derivation of the relevant collision integrals using an unconventional method based on a systematic expansion in powers of connected equal-time 
correlations.~\cite{Fricke97,Hahn21} 


\section{Magnetoelastic bosons in YIG}

\label{sec:bosons}

\subsection{Magnons}

\label{sec:magnons}

\begin{figure}[tb]
\centering
\includegraphics[width=\linewidth]{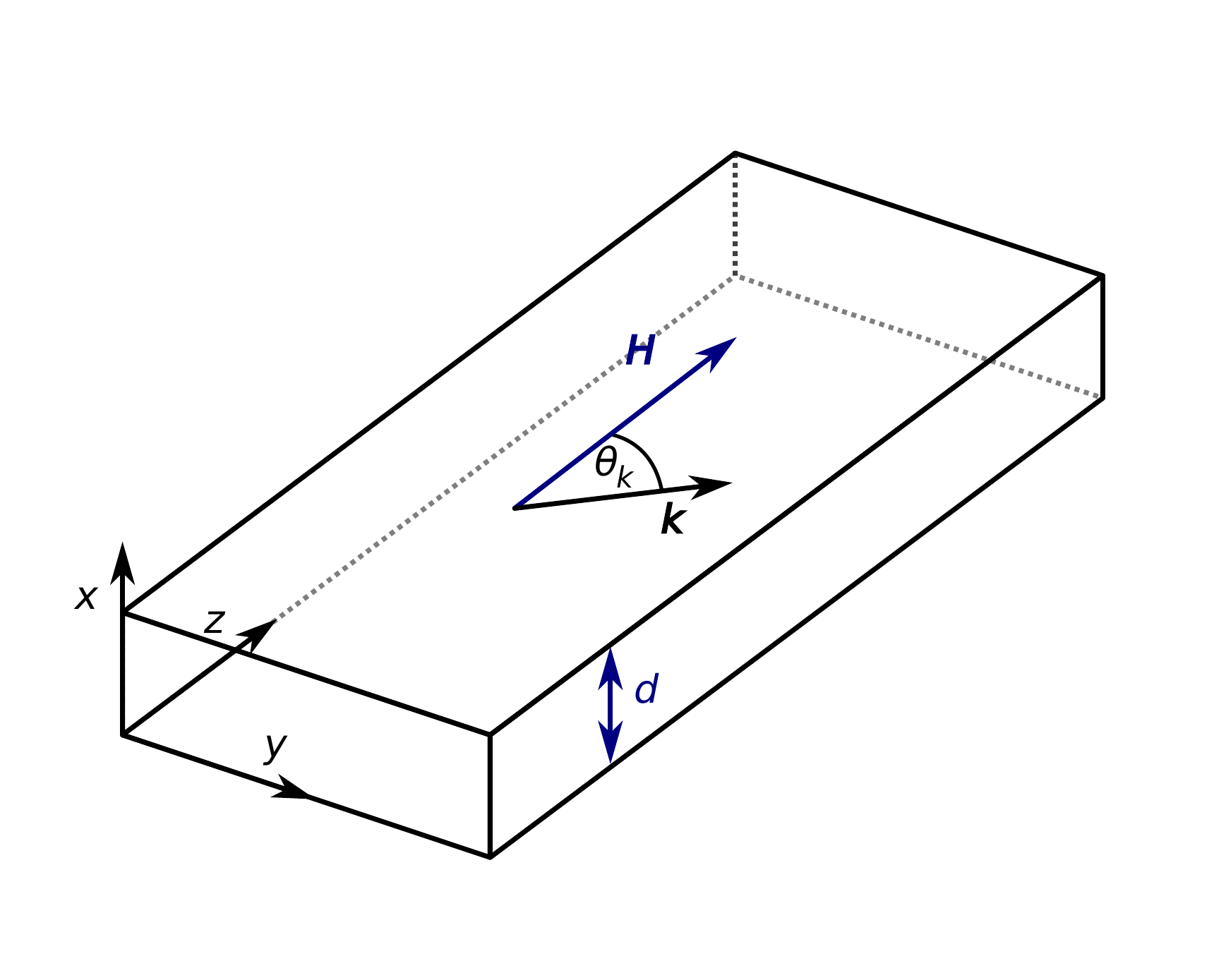}
\caption{
Geometry of a thin YIG film with thickness $d$
in the presence of a uniform external magnetic field $\bm{H} = H \bm{e}_z$ 
parallel to the plane of the film.
In this work we consider only  the lowest,
uniform thickness modes
with in-plane wave vectors 
$ \kv = \bm{e}_z | \kv | \cos \theta_\kv + \bm{e}_y | \kv | \sin \theta_\kv $.
}
\label{fig1}
\end{figure}

\begin{figure}[tb]
\centering
\includegraphics[width=\linewidth,clip=true,trim=100pt 240pt 130pt 260pt]{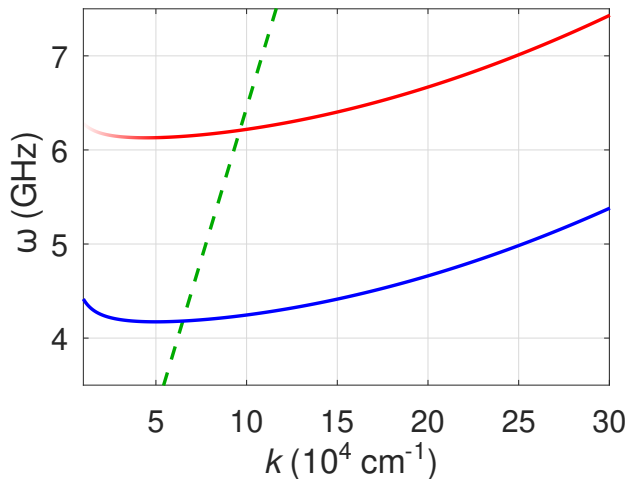}
\caption{
Magnon dispersion $\epsilon_\kv$ as a function of 
wave vector $k = |\kv|$
of a  YIG film
with  thickness $d = 6.7 \, \mu {\rm m}$ in a  magnetic field  $H = 145 \, {\rm mT}$
for momenta parallel (blue) and perpendicular (red) to the magnetic field.
Note that for very small momenta perpendicular to the field,
the magnon dispersion relation \eqref{eq:magnon_disp}
exhibits an unphysical negative slope,
which we indicate by the shading of the red curve \cite{footnote_DE}.
We also  show the dispersion
$\omega_{\kv \bot} = c_{\bot} | \kv |$ 
(dashed green) of the  transverse acoustic phonon mode in YIG. 
}
\label{fig2}
\end{figure}

At room temperature,
the low-energy magnetic properties of YIG can be described by the following effective quantum spin Hamiltonian,~\cite{Cherepanov93,Tupitsyn08,Kreisel09}
\begin{equation} \label{eq:H_eff}
\mathcal{H}_m 
= -\frac{1}{2} \sum_{ij}\sum_{\alpha\beta }\left( J_{ij} \delta^{\alpha\beta} + D_{ij}^{\alpha\beta} \right) S_i^\alpha S_j^\beta - h \sum_i S_i^z ,
\end{equation}
where the indices $i,j=1,\dots,N$ label the sites $\bm{R}_i$ and $\bm{R}_j$ 
of a simple cubic lattice with spacing $a \approx 12.376 \, \mathrm{\r{A}}$,
and $\alpha,\beta \in \{ x,\, y,\, z \}$ 
denote the Cartesian components of the spin operators $\bm{S}_i$ localized at
 lattice sites $\bm{R}_i$.
The ferromagnetic exchange coupling $J_{ij}=J\left(\bm{R}_i-\bm{R}_j\right)$ has the value $J \approx 3.19 \, {\rm K}$ 
if the lattice sites $\bm{R}_i$ and $\bm{R}_j$ are nearest neighbors and vanishes otherwise. Finally, the dipole-dipole interaction tensor is
\begin{equation}
D_{i j}^{\alpha\beta} 
= \left( 1 - \delta_{ i j } \right) \frac{ \mu^2 }{ \left| \bm{R}_{i j} \right|^3 } 
\left[ 3 \hat{{R}}_{i j}^\alpha \hat{{R}}_{i j}^\beta  -\delta^{\alpha\beta} \right] ,
\end{equation}
where $\bm{R}_{ij}=\bm{R}_i-\bm{R}_j$ and  
$\hat{\bm{R}}_{ij}=\bm{R}_{ij}/\left|\bm{R}_{ij}\right|$ is the corresponding unit vector.
The magnetic moment is denoted by $\mu = 2 \mu_B$ where $\mu_B$ is the Bohr magneton. The external magnetic field $ \bm{H} = H \bm{e}_z$ is assumed to be applied in $z$-direction
(the classical ground state is then a saturated ferromagnet with macroscopic magnetization also
pointing in $z$-direction)
and we denote by
$h = \mu H$ the corresponding Zeeman energy.
Having fixed $\mu$ as described above,
we may use the  value of the room-temperature saturation magnetization 
$ 4 \pi M_S = 1750 \, {\rm G}$ of YIG
to determine the effective spin $S = M_s a^3 / \mu \approx 14.2$ of our 
spin model~\cite{Tupitsyn08,Kreisel09}.
This large value of $S$ allows us to bosonize the spin Hamiltonian \eqref{eq:H_eff} via the Holstein-Primakoff transformation~\cite{Holstein40} 
and expand the resulting effective boson Hamiltonian 
with respect to the small parameter $1/S$.
As described in Appendix~\ref{app:magnon},
the quadratic part of the bosonic Hamiltonian is then diagonalized 
by transforming to momentum space and a canonical (Bogoliubov) transformation.
Dropping unimportant constants,
this procedure yields the following quadratic magnon Hamiltonian for YIG,
\begin{equation} \label{eq:H2m}
\mathcal{H}_m^{(2)} = \sum_\kv \epsilon_\kv b^\dagger_\kv b_\kv ,
\end{equation}
where $b^\dagger_\kv$ creates a magnon with momentum 
$\kv$ and energy dispersion $\epsilon_\kv$.
In the thin film geometry shown in Fig.~\ref{fig1}, 
it is sufficient to work with an effective two-dimensional model in order to describe the lowest magnon band of YIG.
Then the long-wavelength dispersion is well approximated by \cite{Kreisel09,Tupitsyn08,Kalinkos86,Hillebrands90}
\begin{align} 
\epsilon_\kv  
=& 
\left[ h + \rho_s \kv^2 + \left( 1- f_\kv \right) \Delta \sin^2\theta_\kv \right]^{1/2}
\nonumber\\
& \times
\left[ h + \rho_s \kv^2 + f_\kv \Delta \right]^{1/2}  .
\label{eq:magnon_disp}
\end{align}
Here,
$\theta_\kv$ is the angle between the magnetic field and the wave vector $\kv$
(see Fig.~\ref{fig1}),
$\rho_s = J S a^2$ and $\Delta = 4\pi M_s$ are the spin stiffness and the dipolar energy scale respectively,
and 
 \begin{equation}
f_\kv = \frac{ 1 - e^{ - | \kv | d } }{   | \kv | d }
 \end{equation}
 is the form factor for a film of thickness $d$.
The magnon dispersion \eqref{eq:magnon_disp} is shown in Fig.~\ref{fig2}
for momenta parallel and perpendicular to the magnetic field 
and experimentally relevant parameters.
Note that in the long-wavelength regime 
probed by experiments~\cite{Bozhko17,Frey21} 
which we aim to describe,
the magnon dispersion \eqref{eq:magnon_disp} of YIG is rather flat.
As a  consequence, all decay processes which do not conserve 
the number of participating magnons are forbidden by energy conservation.
Thus, there is an (approximate) $U(1)$ symmetry for low-energy magnons in YIG, 
which is one of the reasons that magnon condensation is possible in the first place.
Therefore we retain only the number-conserving two-body  magnon-magnon interaction
\begin{eqnarray} \label{eq:Hm4}
\mathcal{H}_m^{(4)} & = &  
\frac{1}{N} \sum_{\kv_1 \dots \kv_4} \delta_{\kv_1+\dots+\kv_4, 0} 
\frac{1}{ 4 } U_{\kv_1, \kv_2; \kv_3, \kv_4} 
 \nonumber
 \\
 & & \times
b^\dagger_{- \kv_1} b^\dagger_{-\kv_2} b_{\kv_3} b_{\kv_4} , 
\end{eqnarray}
where the interaction vertex $U_{ \kv_1, \kv_2; \kv_3, \kv_4}$ is explicitly 
given in Eq.~\eqref{eq:vertex4} of Appendix~\ref{app:magnon}.
 
\subsection{Phonons and magnetoelastic hybridization in YIG}

So far,
we have considered only the magnon subsystem.
In order to address the accumulation of magnetoelastic bosons,
we should also take the  phonons in YIG into account.
At long wavelengths,
the three relevant acoustic phonon branches of YIG are described by
the following quadratic phonon Hamiltonian,
\begin{equation} \label{eq:H2e}
\mathcal{H}_p^{(2)} = 
\sum_{ \kv \lambda } \omega_{\kv\lambda} \left( a_{\kv\lambda}^\dagger a_{\kv\lambda}+\frac12 \right) ,
\end{equation}
where $a_{\kv\lambda}^\dagger$ creates 
a phonon with momentum $\kv$, polarization $\lambda$, 
and energy $\omega_{\kv\lambda} = c_\lambda |\kv|$, where $c_{\lambda}$
are the phonon velocities.
It is well known \cite{Gurevich96,Gilleo58} that in YIG
there are two degenerate transverse ($\lambda = \bot 1 ,\, \bot 2$) phonon
modes
with phonon velocity $c_{\bot 1} = c_{\bot 2} \equiv c_\bot = 3.843\times10^5 \, \mathrm{cm}/\mathrm{s}$,
and one longitudinal ($\lambda = \|$) mode 
with  velocity $c_\| = 7.209\times10^5 \, \mathrm{cm}/\mathrm{s}$.
Interactions between the phonons can be safely ignored because of the 
large mass density 
$\rho = 5.17 \, \mathrm{g}/\mathrm{cm}^3$ of YIG.~\cite{Gurevich96}
The transverse phonon dispersion $\omega_{\kv  \bot}$ is shown
in Fig.~\ref{fig2}  as a dashed green line.

The coupling between the magnons and the phonons 
arises both from the dependence of the exchange interaction on the ionic positions
as well as from relativistic effects involving the charge degrees of freedom which
cannot be taken into account directly  within  an effective spin model.
As the latter is usually dominant in collinear magnets 
at low energies,\cite{Gurevich96}
we opt to derive the magnon-phonon interactions by 
quantizing the  phenomenological expression 
for the classical magnetoelastic energy. This strategy was pioneered by
Abrahams and Kittel \cite{Abrahams52} and more recently adopted in
Ref.~[\onlinecite{Rueckriegel14}].
At long wavelengths,
the relevant contribution to the classical magnetoelastic energy is
\begin{equation} \label{eq:E_me}
E_{\rm me} = \frac{ n }{ M_s^2 } \int d^3 r \sum_{ \alpha \beta } B^{ \alpha \beta } M^\alpha ( \bm{r} ) M^\beta ( \bm{r} ) X^{ \alpha \beta } ( \bm{r} ) ,
\end{equation}
where 
$\bm{M} ( \bm{r} )$ is the local magnetization,
$X^{ \alpha \beta } ( \bm{r} )$ is the symmetric strain tensor,
$n = a^{ - 3 }$ is the number density of magnetic ions,
and $B^{ \alpha \beta }$ are phenomenological magnetoelastic constants.
For a cubic lattice,
these constants can be written as
$B^{ \alpha \beta } = \delta^{ \alpha \beta } B_\parallel + ( 1 - \delta^{ \alpha \beta } ) B_\bot$,
where $B_\| = 47.8 \,\mathrm{K}$ and $B_\bot = 95.6 \, \mathrm{K}$ for YIG~\cite{Gurevich96,Eggers63,Hansen73}.
The magnetoelastic energy (\ref{eq:E_me}) can then be quantized by replacing 
$\bm{M}( \bm{r} = \bm{R}_i ) \to \mu n \bm{S}_i$
and expanding the strain tensor $X^{ \alpha \beta } ( \bm{r} )$ 
in terms of the phonon operators $a_{\kv\lambda}$ and $a_{\kv\lambda}^\dagger$.
This procedure,
outlined in Appendix.~\ref{app:phonon} 
and discussed in detail in Ref.~\onlinecite{Rueckriegel14},
yields to lowest order in $1/S$ the following Hamiltonian for the hybridization of magnons and phonons,
\begin{equation} \label{eq:H2_mp}
\mathcal{H}_{mp}^{(2)} 
= 
\frac{1}{2} \sum_{ \kv \lambda} \Gamma_{ \kv \lambda }
\left( 
a_{ -\kv \lambda } + a_{ \kv \lambda }^\dagger
\right) b_\kv
+ {\rm h.c.} ,
\end{equation}
where h.c.~denotes the hermitian conjugate,
and the hybridization vertices $\Gamma_{\kv \lambda}$ are given explicitly 
in Eqs.~\eqref{eq:gamma} and \eqref{eq:Gamma} of Appendix~\ref{app:phonon}.
Higher order magnon-phonon interactions open up additional decay 
channels.\cite{Rueckriegel14,Streib19}
However, 
for YIG films the contribution of these processes is generally several orders of 
magnitude smaller 
than the contribution of the magnon-magnon interaction (\ref{eq:Hm4}) 
at long wavelengths \cite{Demokritov06,Rueckriegel14,Streib19},
which justifies neglecting them.

\begin{figure}[tb]
\centering
\includegraphics[width=\linewidth,clip=true,trim=100pt 240pt 130pt 260pt]{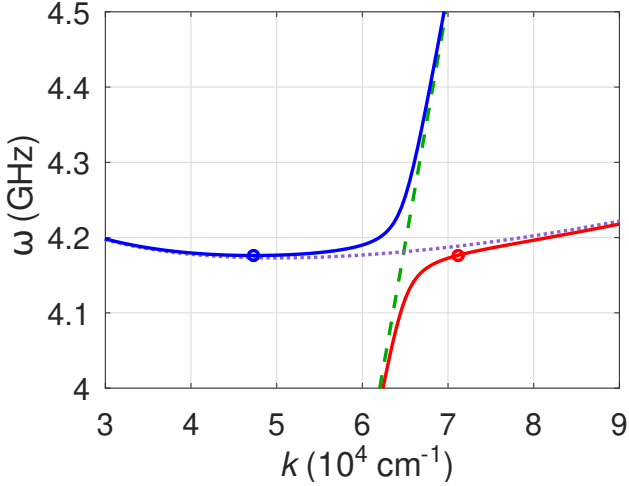}
\caption{
Dispersions of magnetoelastic modes in YIG for momenta 
parallel to the magnetic field:   the blue solid curve represents the $+$ branch while the red solid curve represents the $-$ branch. The blue circle shows the minimum of the dispersion of the $+$ branch and the red circle marks the point where the dispersion of the $-$ branch has the same value. The magnon and phonon dispersions in absence of hybridization are also shown as dotted purple and dashed green line respectively.
The film thickness is chosen as $d = 6.7 \, \mu {\rm m}$ and the magnetic field strength is $H = 145 \, {\rm mT}.$
}
\label{fig3}
\end{figure}

In the following,
we will focus solely on the transverse phonon branches and 
drop the longitudinal ones,
because in thin YIG films only the two transverse branches hybridize with the magnons
in the experimentally relevant region \cite{Frey21,Rueckriegel14}.
To describe the magnetoelastic modes,
we may furthermore neglect the non-resonant terms 
$a_{ -\kv \lambda } b_\kv$ and $a_{ -\kv \lambda }^\dagger b_\kv^\dagger$
in the hybridization Hamiltonian (\ref{eq:H2_mp}) as discussed in
Refs.~[\onlinecite{Takahashi16,Bozhko17}].
In this approximation,
the quadratic Hamiltonian 
 \begin{equation}
{\cal H}^{(2)} = {\cal H}_{m}^{(2)} + {\cal H}_{p}^{(2)} + {\cal H}_{mp}^{(2)}
 \end{equation}
of the coupled magnon-phonon system can be diagonalized by the unitary transformation
\begin{equation}
\left( \begin{matrix}
b_\kv \\ a_{ \kv \bot 1 } \\ a_{ \kv \bot 2 }
\end{matrix} \right)
=
\left( \begin{matrix} 
\bm{\phi}_{ \kv + } , \bm{\phi}_{ \kv - } , \bm{\phi}_{ \kv p }
\end{matrix} \right)
\left( \begin{matrix}
\psi_{ \kv + } \\ \psi_{ \kv - } \\ \psi_{ \kv p } 
\end{matrix} \right) .
\end{equation}
Here, $\psi_{ \kv +}$, $\psi_{\kv -}$ and $\psi_{\kv p}$ are canonical bosonic 
annihilation operators associated with magnetoelastic modes, and the
three column vectors $\bm{\phi}_{\kv +}$, $\bm{\phi}_{\kv -}$,
$\bm{\phi}_{\kv p}$ are given by
\begin{subequations}
\begin{align}
\bm{\phi}_{ \kv \pm } 
=& 
\frac{ 
\left( 2 \left( E_{ \kv \pm } - \omega_{ \kv \bot } \right) , \Gamma_{ \kv \bot 1 } , \Gamma_{ \kv \bot 2 } \right)^T
}{ 
\sqrt{ 4 \left( E_{ \kv \pm } - \omega_{ \kv \bot } \right)^2 
+ \left| \Gamma_{ \kv \bot 1 } \right|^2 + \left| \Gamma_{ \kv \bot 2 } \right|^2 } 
}
, 
\label{eq:MEM_ev}
\\
\bm{\phi}_{ \kv p } 
=& 
\frac{
\left( 0 , - \Gamma_{ \kv \bot 2 }^* , \Gamma_{ \kv \bot 1 }^* \right)^T
}{
\sqrt{ \left| \Gamma_{ \kv \bot 1 } \right|^2 + \left| \Gamma_{ \kv \bot 2 } \right|^2 }
} .
\end{align}
\end{subequations}
These vectors can  be identified with eigenvectors of the relevant $3 \times 3$ 
Hamiltonian matrix.
The dispersions of the two magnetoelastic modes are given by
\begin{align} 
E_ { \kv \pm } 
=& 
\frac{1}{2} \left[
\epsilon_\kv + \omega_{ \kv \bot }
\vphantom{ \sqrt{ \left( \epsilon_\kv - \omega_{ \kv \bot } \right)^2 
+ \left| \Gamma_{ \kv \bot 1 } \right|^2 + \left| \Gamma_{ \kv \bot 2 } \right|^2 }  }
\right.
\nonumber\\
& \left. \phantom{ \frac{1}{2} }
\pm \sqrt{ \left( \epsilon_\kv - \omega_{ \kv \bot } \right)^2 
+ \left| \Gamma_{ \kv \bot 1 } \right|^2 + \left| \Gamma_{ \kv \bot 2 } \right|^2 } 
\right].
\label{eq:E_pm}
\end{align}
In Fig.~\ref{fig3} a graph of these dispersions is shown
for a YIG film with experimentally relevant parameters.
The diagonalized quadratic Hamiltonian of the coupled magnon-phonon system then takes the simple form
\begin{align} \label{eq:H2}
{\cal H}^{ (2) } =& \sum_\kv \Bigl[
E_{ \kv + } \psi_{ \kv + }^\dagger \psi_{ \kv + } +
E_{ \kv - } \psi_{ \kv - }^\dagger \psi_{ \kv - } 
\nonumber\\
& \phantom{ \sum_\kv }
+
\omega_{ \kv \bot } \psi_{ \kv p }^\dagger \psi_{ \kv p }
\Bigr].
\end{align}
The purely phononic operators $\psi_{ \kv p }$ will not play a role in the following and are hence discarded.
Expressing the quartic magnon-magnon interaction (\ref{eq:Hm4})
in terms of the creation and annihilation operators $\psi_{ \kv \pm }^\dagger$ and $\psi_{ \kv \pm }$ of the magnetoelastic bosons and dropping some subleading intermodal terms (see below),
we obtain
\begin{align} 
\mathcal{H}_m^{(4)} 
 \approx & 
\frac{1}{N} \sum_{\kv_1 \dots \kv_4} \delta_{\kv_1+\dots+\kv_4, 0} 
\frac{1}{ 4 } 
\nonumber\\
&
\times \Bigl[
U_{1, 2; 3, 4}^{ + + + + } \psi^\dagger_{ -1 + } \psi^\dagger_{ -2 + } \psi_{ 3 + } \psi_{ 4 + }
\nonumber\\
& \phantom{ \times }
+ 
U_{1, 2; 3, 4}^{ - - - - } \psi^\dagger_{ -1 - } \psi^\dagger_{ -2 - } \psi_{ 3 - } \psi_{ 4 - } 
\nonumber\\
& \phantom{ \times }
+ 
U_{1, 2; 3, 4}^{ + + - - } \psi^\dagger_{ -1 + } \psi^\dagger_{ -2 + } \psi_{ 3 - } \psi_{ 4 - } 
\nonumber\\
& \phantom{ \times }
+ 
U_{1, 2; 3, 4}^{ - - + + } \psi^\dagger_{ -1 - } \psi^\dagger_{ -2 - } \psi_{ 3 + } \psi_{ 4 + } 
\Bigr] ,
\label{eq:Hme4}
\end{align}
where the subscripts $1,2, \ldots$ represent $\kv_1, \kv_2, \ldots$. The
interaction vertices are 
\begin{subequations}
 \label{eq:U4def}
\begin{align}
U_{1, 2; 3, 4}^{ \pm \pm\pm \pm } 
=& \left( \phi_{ -1 \pm }^1 \right)^* \left( \phi_{ -2 \pm }^1 \right)^* 
\phi_{ 3 \pm }^1 \phi_{ 4 \pm }^1 U_{1, 2; 3, 4} , \\
U_{1, 2; 3, 4}^{ \pm \pm \mp \mp } 
=& \left( \phi_{ -1 \pm }^1 \right)^* \left( \phi_{ -2 \pm }^1 \right)^* 
\phi_{ 3 \mp }^1 \phi_{ 4 \mp }^1 U_{1, 2; 3, 4} ,
\end{align}
\end{subequations}
and $\phi_{ \kv \pm }^1 = (1,0,0) \cdot \bm{\phi}_{ \kv \pm }$
are the magnonic components 
of the magnetoelastic wave functions defined in Eq.~\eqref{eq:MEM_ev}.
The vertices 
$U_{1, 2; 3, 4}^{ \pm \pm\pm \pm } $
describe $++ \Leftrightarrow ++$ and $-- \Leftrightarrow --$
intramodal scattering events where
the number of each magnetoelastic boson is conserved.
They are responsible for the rapid thermalization of the pumped magnon gas 
away from the hybridization area.
The other class of
vertices
$U_{1, 2; 3, 4}^{ \pm \pm \mp \mp } $ describe
$++ \Leftrightarrow --$
intermodal scattering events where
the number of magnetoelastic
bosons in the $+$ or $-$ branch changes by two 
while the total
number of magnetoelastic bosons is conserved.
Consequently, 
these processes exchange both energy and particles between the two magnetoelastic modes and thus are important for the thermalization of the low- and high-energy parts of the magnon spectrum.
Because of their energy and momentum conservation constraints, 
they furthermore lead to a direct coupling of the region around the bottom of the magnon dispersion on the $+$ mode and the nearly degenerate hybridization area of the $-$ mode.
Hence,
we expect these intermodal processes to be crucial for the eventual appearance of a magnetoelastic accumulation. 

Note that in Eq.~\eqref{eq:Hme4}
we have followed the ansatz described in Ref.~[\onlinecite{Bozhko17}] and
dropped two types of subleading intermodal scattering processes: 
A $+- \Leftrightarrow +-$ process 
which does not change the number of bosons in both branches,
as well as $++ \Leftrightarrow +-$ and $-- \Leftrightarrow -+$ processes 
where the number of bosons on each branch changes only by one.
While these scattering processes give rise to additional thermalization channels,
we do not expect them to substantially affect the steady state. 
The first process only redistributes the bosons within the two branches,
similar to the intramodal scattering.
On the other hand, 
the second process can lead to an exchange of bosons between the bottom of the $+$ mode and the energetically degenerate hybridization area of the $-$ mode.
However, 
to satisfy energy and momentum conservation,
such a scattering requires the participation of high-energy magnons from the $-$ branch. 
Close to the steady state, 
we generally expect such processes that also involve high-energy bosons
to be less important than the direct scattering between the low-energy bosons
and the macroscopically occupied condensate.

\section{Accumulation of magnetoelastic bosons}

\label{sec:Accumulation}

In order to describe the experimentally observed accumulation of magnetoelastic 
bosons,~\cite{Bozhko17,Frey21}
we derive in this section quantum kinetic equations for the single-particle distribution functions 
and the condensation amplitudes  of the magnetoelastic modes associated with the
bosonic operators $\psi_{ \kv \pm }$.
The kinetic equations are then solved self-consistently to obtain a non-equilibrium steady 
state which can be  compared with experiments.

\subsection{Quantum kinetic equations}

The dynamics of the connected single-particle distribution function
of the magnetoelastic modes,
\begin{equation}
n_{ \kv \pm } 
\equiv \langle \psi_{ \kv \pm }^\dagger \psi_{ \kv \pm } \rangle^c
\equiv \langle \psi_{ \kv \pm }^\dagger \psi_{ \kv \pm } \rangle - \left| \Psi_{ \kv \pm } \right|^2 ,
\end{equation}
and the dynamics of the associated  condensate amplitude (vacuum expectation value)
\begin{equation}
\Psi_{ \kv \pm } \equiv \langle \psi_{ \kv \pm } \rangle 
\end{equation}
can be obtained from the Heisenberg equations of motion of the 
Bose operators $\psi_{ \kv \pm }$. We write the 
equation of motion for the single-particle distribution function in the form
\begin{equation} \label{eq:dynamics_n}
\partial_t n_{ \kv \pm } = I_{ \kv \pm } ,
\end{equation}
where $I_{\kv \pm}$ is the relevant collision integral. 
The derivation of this collision integral is outlined in Appendix~\ref{app:collision} and the approximate expression 
sufficient for our purpose is given below in Eq.~\eqref{eq:coll_inta}.
The equation of motion for the condensate amplitude is
\begin{widetext}
\begin{equation} \label{eq:dynamics_Psi}
\partial_t \Psi_{ \kv \pm } + i \left( E_{ \kv \pm } - \mu_c \right) \Psi_{ \kv \pm }
+ \frac{ i }{ 2 N } \sum_{ 1 2 3 } \delta_{ 1 + 2 + 3 , \bm{k} }
\left[ 
U_{ - \bm{k} , 1 ; 2 , 3 }^{ + + + + } \Psi_{ -1 + }^* \Psi_{ 2 + } \Psi_{ 3 + } +
U_{ - \bm{k} , 1 ; 2 , 3 }^{ + + - - } \Psi_{ -1 + }^* \Psi_{ 2 - } \Psi_{ 3 - }  
\right]
= \tilde{I}_{ \kv \pm } ,
\end{equation}
\end{widetext}
where 
$\mu_c$ is the chemical potential of the condensate and
the collision integral $\tilde{I}_{ \kv \pm }$ describes scattering into and out of the condensate. 
The approximate expression for this collision integral that we use 
is given in Eq.~\eqref{eq:coll_intb} below;
for more details we refer to Appendix~\ref{app:collision}.
For a realistic description of the experimental setup,
this chemical potential $\mu_c$ of the condensate is necessary to
take into account the approximate number conservation of the magnon subsystem.
Physically, the finite value of  $\mu_c$ is generated by the external pumping
and  is one of the parameters which  characterize 
the non-equilibrium steady state.
The collision integrals  $I_{ \kv \pm }$ and $\tilde{I}_{ \kv \pm }$ 
on the right-hand sides of the equations of 
motion \eqref{eq:dynamics_n} and \eqref{eq:dynamics_Psi}
describe the effect
of  the quartic interaction (\ref{eq:Hme4}) on the dynamics;
in general, $I_{ \kv \pm }$ and $\tilde{I}_{ \kv \pm }$ are complicated functionals
of higher-order connected correlation functions, which satisfy additional equations of motion
involving even higher-order correlation functions.
One of the central problems in quantum kinetic theory is to find a 
good truncation strategy of this 
infinite hierarchy of equations of motion. Here we us the method of expansion
in connected equal-time correlations developed in Ref.~[\onlinecite{Fricke97}] which
two of us have recently used \cite{Hahn21}
to develop a microscopic description of the effect of magnon decays on parametric pumping 
of magnons in YIG. An advantage of this method is that 
it directly produces equal-time correlations and that it offers a systematic
truncation strategy in powers of connected correlations.
The dominant contributions to the collision integrals
$I_{ \kv \pm }$ and $\tilde{I}_{ \kv \pm }$ in
Eqs.~\eqref{eq:dynamics_n} and \eqref{eq:dynamics_Psi}
are given in Appendix~\ref{app:collision}, 
where we also
give a diagrammatic representation of the various terms contributing to
 $I_{ \kv \pm }$ and $\tilde{I}_{ \kv \pm }$.
Because the magnon-magnon interaction (\ref{eq:Hme4}) in YIG is suppressed 
by the small factor of $1/S$,
for our purpose it is sufficient to truncate the hierarchy of equations of motion at second order in the interaction.
This yields the following expressions for the collision integrals
on the right-hand sides of the equations of motion \eqref{eq:dynamics_n} and \eqref{eq:dynamics_Psi}:
\begin{widetext}
%
%
\begin{align}
I_{ \kv \pm } =
&
\frac{ \pi }{ 4 N^2 } \sum_{ 1 2 3 } \delta_{ \kv + 1 , 2 + 3 } 
\sum_{ r = \pm } \left| U_{ -\kv , -1 ; 2 , 3 }^{ \pm \pm r r } \right|^2
\nonumber\\
&
\times \Bigl\{
\delta \left( E_{ \kv \pm } + E_{ 1 \pm } - E_{ 2 r } - E_{ 3 r } \right)
\bigl[
\left( 1 + n_{ \kv \pm } \right) \left( 1 + n_{ 1 \pm } \right) n_{ 2 r } n_{ 3 r } -
n_{ \kv \pm } n_{ 1 \pm } \left( 1 + n_{ 2 r } \right) \left( 1 + n_{ 3 r } \right)
\bigr]
\nonumber\\
& \phantom{ \times a }
- \frac{ 1 }{ 2 }
\delta \left( E_{ \kv \pm } + \mu_c - E_{ 2 r } - E_{ 3 r } \right)
 \left| \Psi_{ 1 \pm } \right|^2
\bigl[
\left( 1 + n_{ \kv \pm } \right)  n_{ 2 r } n_{ 3 r } -
n_{ \kv \pm } \left( 1 + n_{ 2 r } \right) \left( 1 + n_{ 3 r } \right)
\bigr]
\nonumber\\
& \phantom{ \times a }
+ 
\delta \left( E_{ \kv \pm } + E_{ 1 \pm } - E_{ 2 r } - \mu_c \right) 
\left| \Psi_{ 3 r } \right|^2
\bigl[
\left( 1 + n_{ \kv \pm } \right) \left( 1 + n_{ 1 \pm } \right)  n_{ 2 r }  -
n_{ \kv \pm } n_{ 1 \pm } \left( 1 + n_{ 2 r } \right) 
\bigr]
\Bigr\} ,
 \label{eq:coll_inta}
\\[.25cm]
\tilde{ I }_{ \kv \pm } =
&
\frac{ \pi }{ 8 N^2 } \Psi_{ \kv \pm } \sum_{ 1 2 3 } \delta_{ \kv + 1 , 2 + 3 } 
\sum_{ r = \pm } \left| U_{ -\kv , -1 ; 2 , 3 }^{ \pm \pm r r } \right|^2
\nonumber\\
&
\times 
\delta \left( E_{ 1 \pm } + \mu_c - E_{ 2 r } - E_{ 3 r } \right)
\bigl[
n_{ 1 \pm } \left( 1 + n_{ 2 r } \right) \left( 1 + n_{ 3 r } \right) -
\left( 1 + n_{ 1 \pm } \right) n_{ 2 r } n_{ 3 r } 
\bigr]
.
 \label{eq:coll_intb}
\end{align}
%
%
\end{widetext}
Note that apart from the additional $\pm$ mode label, 
the resulting kinetic equations coincide with the standard Boltzmann equations 
for Bose gases known from the literature \cite{Zaremba99}.

\subsection{Non-equilibrium steady state}

In principle,
it would be desirable to directly simulate the temporal evolution of the distribution functions and condensate amplitudes that is generated by the coupled integro-differential equations \eqref{eq:dynamics_n} and \eqref{eq:dynamics_Psi}
with the collision integrals given by Eqs.~\eqref{eq:coll_inta} and (\ref{eq:coll_intb}).
However,
this is a computationally very demanding task because it requires us to cover a large region of momentum space up to comparatively large energies so that thermalization can occur, 
while at the same time a very fine momentum resolution is necessary to resolve the bottom of the magnon spectrum as well as the energetically degenerate hybridization area in sufficient detail.
To circumvent these computational difficulties,
we focus on the steady state that eventually forms in the parametrically pumped magnon gas.
Then we can take advantage of the fact that
the magnon-magnon interaction \eqref{eq:Hm4} efficiently thermalizes
the magnon gas to a quasi-equilibrium steady state 
characterized by a finite chemical potential $\mu_m$.
When  this chemical potential approaches the minimum of the magnon dispersion,
a condensate is formed.\cite{Demokritov06,Demidov07,Dzyapko07,Demidov08a,Demokritov08,Demidov08b,Serga14,Clausen15a,Clausen15b}
If the pumping is turned off,
the chemical potential and the condensate slowly decay 
on time scales governed by the weak magnon-phonon interactions.\cite{Demokritov06,Clausen15a,Clausen15b}
As the magnon-phonon hybridization which we aim to include 
only affects the mode dispersions and interaction amplitudes
in a tiny region of momentum space,
we may assume that the magnon gas is thermalized almost everywhere in momentum space.
In this case the distribution functions of the magnetoelastic modes are described 
by the incoherent superposition
\begin{equation} \label{eq:superposition}
n_{ \kv \pm } = 
\left| \phi_{ \kv \pm }^1 \right|^2 n_{ \kv m } +
\left( 1 - \left| \phi_{ \kv \pm }^1 \right|^2 \right) n_{ \kv p }
\end{equation}
of the thermalized magnon and phonon distributions
\begin{subequations}
\begin{align}
\label{eq:th_magnon}
n_{ \kv m } =& \frac{ 1 }{ e^{ \left( \epsilon_{ \kv } - \mu_m \right) / T_m } - 1 } , \\
n_{ \kv p } =& \frac{ 1 }{ e^{ \omega_{ \kv \bot }  / T } - 1 } .
\end{align}
\end{subequations}
Here
we take into account that the magnon temperature $T_m$ in the steady state 
can deviate from the  temperature $T$ of the phonons.
Since these distribution functions annihilate the collision integrals~\eqref{eq:coll_inta} and (\ref{eq:coll_intb}) almost everywhere in momentum space,
we can now focus on the small region in momentum space 
where deviations from Eqs.~\eqref{eq:superposition} are expected to occur: 
The hybridization area where magnons and phonons mix, 
and the bottom of the magnon spectrum that is energetically degenerate with the hybridization, see Fig.~\ref{fig3}.
The problem is then reduced to the calculation of
the change 
in the distribution functions and the condensate amplitudes of the two magnetoelastic modes
in these two regions.
To this end, 
we develop a self-consistent solution of the kinetic equations \eqref{eq:dynamics_n} and \eqref{eq:dynamics_Psi} as follows:
in a non-equilibrium steady state,
the distribution functions and the condensate amplitudes are stationary, so that
\begin{subequations}
\begin{align}
\partial_t n_{ \kv \pm } =& 0 , \\
\partial_t \Psi_{ \kv \pm } =& 0 .
\end{align}
\end{subequations}
Starting from an initial guess for $n_{ \kv \pm }$ and $\Psi_{ \kv \pm }$,
we can then use the equations of motion 
\eqref{eq:dynamics_n} and \eqref{eq:dynamics_Psi}
to determine new values for the distribution functions $n_{ \kv \pm }$ 
as well as the condensate amplitudes $\Psi_{ \kv \pm }$.
These are in turn used to determine the new values of the collision integrals \eqref{eq:coll_inta}
and \eqref{eq:coll_intb}.
This procedure is iterated until convergence is achieved.
As initial conditions for the self-consistency loop,
we choose the incoherent superposition \eqref{eq:superposition} for $n_{ \kv \pm }$,
whereas the initial condensate density is estimated as follows.
We neglect the collision integral in the equation of motion of the condensate amplitude \eqref{eq:dynamics_Psi} and set the loop momenta in the Gross-Pitaevskii terms equal the external momenta. Demanding that the time derivative of the condensate amplitude vanishes, we then obtain
\begin{equation} \label{eq:Psi_initial}
\left| \Psi_{ \kv r } \right| 
= \delta_{ \kv , \kv_{\rm min} } \delta_{ r, + }
\sqrt{ N \left| \frac{ \mu_c - E_{ \kv + } }{ U_{ - \kv , - \kv ; \kv , \kv }^{ + + + + } } \right| } .
\end{equation}
Here, 
$\kv_{\rm min}$ denotes the wave vector of the minimum of the magnon dispersion
that is located in the $+$ branch of the magnetoelastic spectrum.
Furthermore,
changes in the distribution of the thermal magnon cloud
are accounted for by also determining the magnon chemical potential $\mu_m$ and temperature $T_m$ 
self-consistently at each iteration.
The phonon temperature $T$ on the other hand is kept fixed,
reflecting the fact that the phonons act as a thermal bath for the magnons.

\begin{figure}[tb]
\centering
\begin{minipage}{16pt}
(a)
\end{minipage}
\begin{minipage}{0.91\linewidth}
\includegraphics[clip=true,trim=10pt 180pt 0pt 200pt,width=\linewidth]{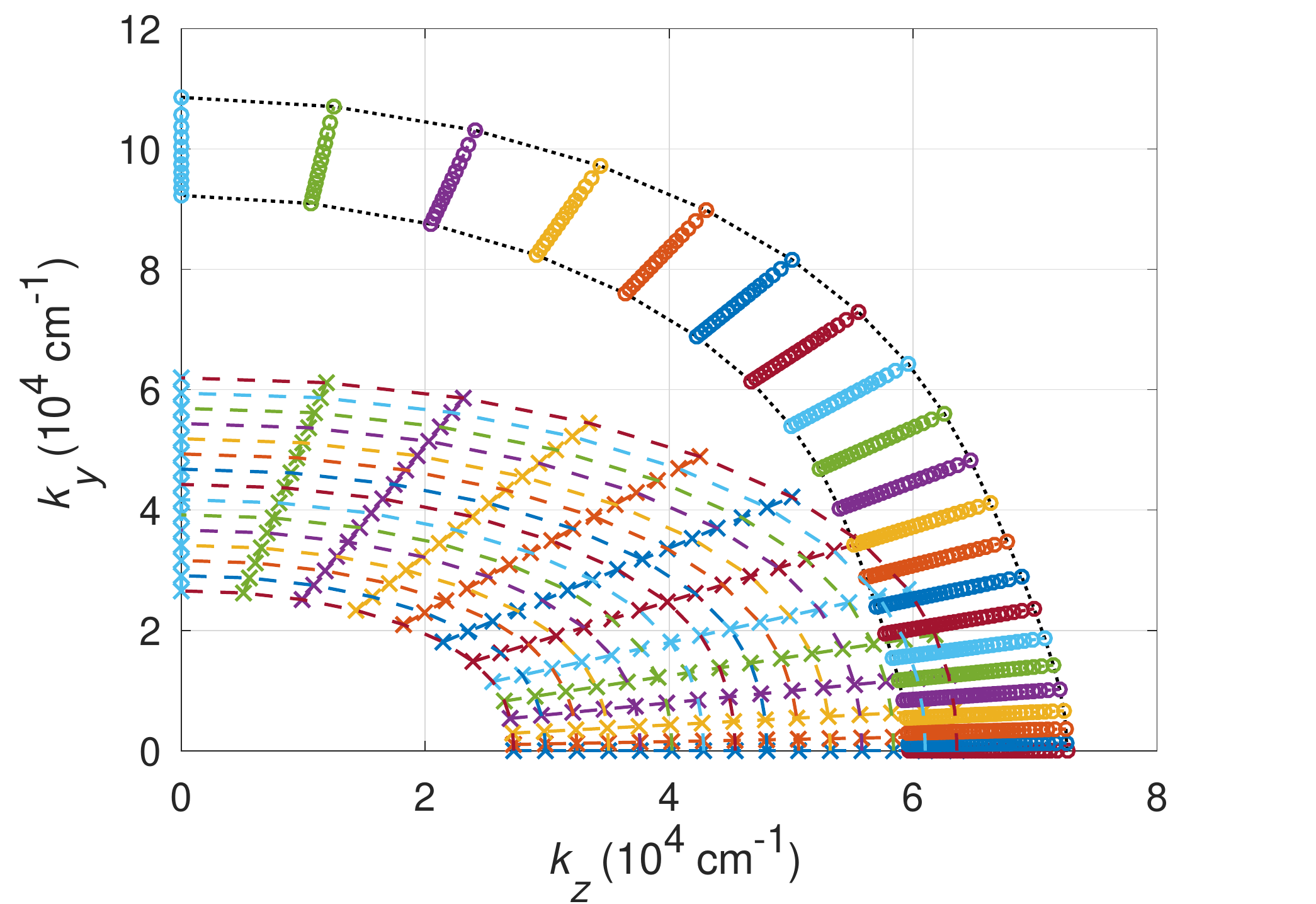}
\end{minipage}
\begin{minipage}{16pt}
(b)
\end{minipage}
\begin{minipage}{0.91\linewidth}
\includegraphics[clip=true,trim=10pt 180pt 0pt 200pt,width=\linewidth]{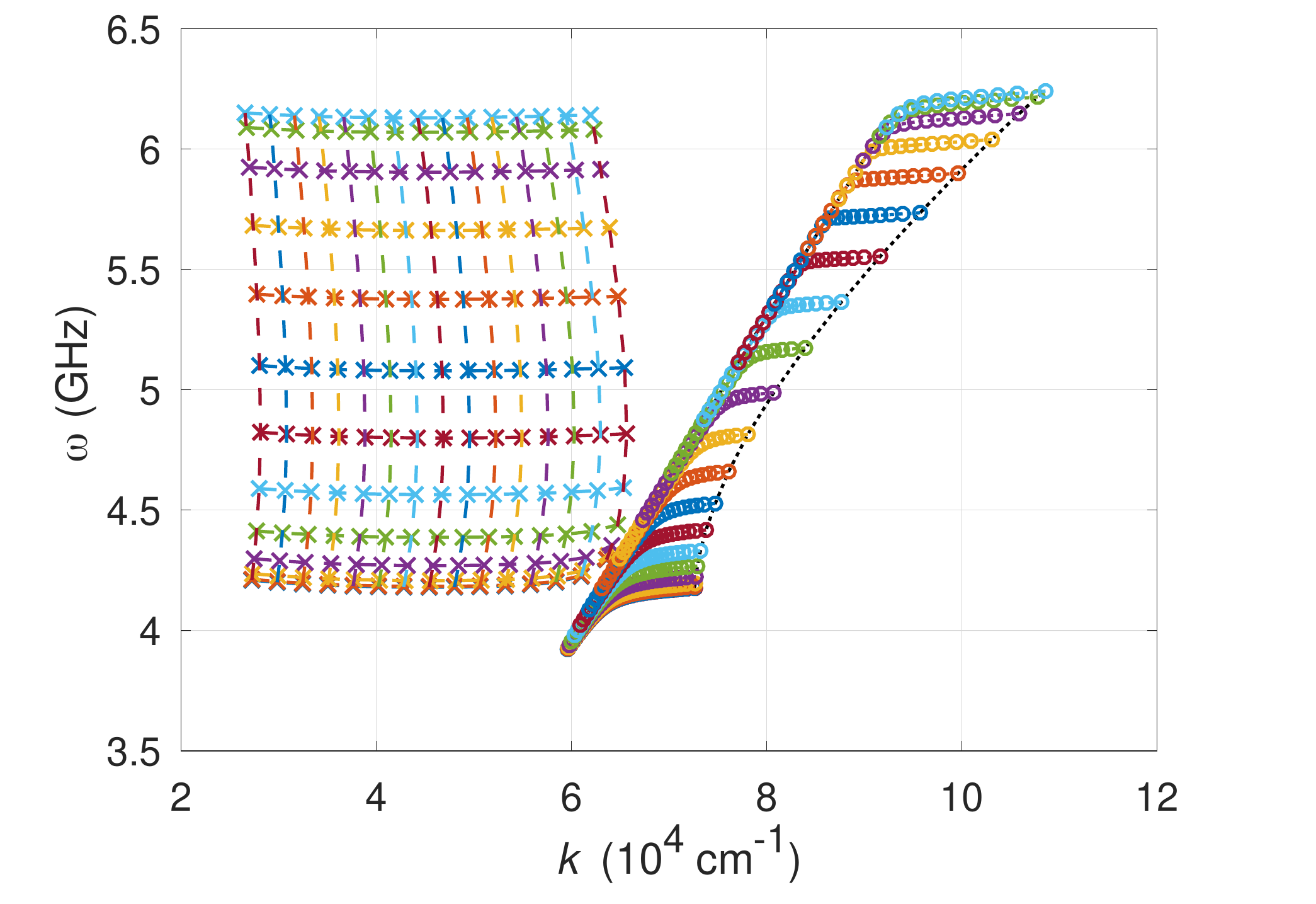}
\end{minipage}
\caption{
The mesh of wave vectors $\kv = \bm{e}_z k_z + \bm{e}_y k_y$ 
consisting of $790$ grid points used
for  the numerical solution of the kinetic equations in this section. 
Crosses denote grid points for the upper ($+$) branch and circles denote grid points for the lower ($-$) branch. 
The upper figure (a)  shows the grid in momentum space while the lower figure 
(b) shows the grid on the plane spanned by $k = | \kv |$ and the
excitation frequency $\omega$
using the same color coding and 
symbols  as in (a).}
\label{fig4}
\end{figure}

For the explicit numerical solution,
we parametrize the wave vectors $\kv$ by choosing
$N_\theta$ angles $\theta_\kv \in [ 0 , \pi/2 ]$ and 
$N_k$ points for different lengths $k = | \kv |$ of the wave vectors.
For each angle $\theta_\kv$,
the $k$-values are chosen such that they are centered around the minimum of the magnon dispersion for the upper ($+$) mode
and the hybridization area for the lower ($-$) mode, see Fig.~\ref{fig3}.
The resulting non-uniform mesh in momentum space is illustrated in Fig.~\ref{fig4}.
All modes outside this mesh are modeled with the quasi-equilibrium distribution \eqref{eq:superposition}. 
To reproduce the experimental situation,
the phonon temperature is fixed at room temperature,
$T = 290 \, {\rm K}$,
while the external magnetic field and the thickness of the YIG film are set to
$ H = 145 \, {\rm mT} $ and $ d = 6. 7\, \mu {\rm m} $ respectively.
The condensate chemical potential is set to 
$ \mu_c = 0.995 \, \epsilon_{ \kv_{\rm min} } $  while 
we use 
$T_m = T$ and $\mu_m = 0.98 \, \epsilon_{ \kv_{\rm min} }$ as
 initial conditions for the self-consistency loop 
of the temperature and chemical potential of the thermal magnons \cite{footnote_muc}. 
The system size appearing in the initial value \eqref{eq:Psi_initial}
for the condensate amplitude is set to $N = 8.0802 \times 10^6$.

\begin{figure}[tb]
\centering
\begin{minipage}{16pt}
(a)
\end{minipage}
\begin{minipage}{0.91\linewidth}
\includegraphics[scale=0.25]{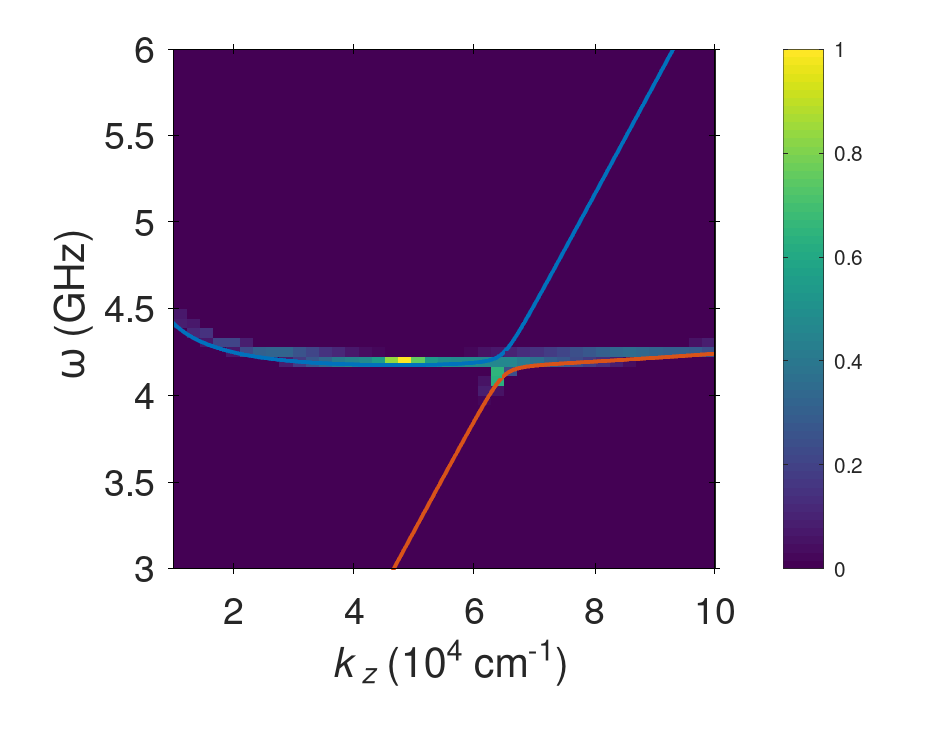}
\end{minipage}
\begin{minipage}{16pt}
(b)
\end{minipage}
\begin{minipage}{0.91\linewidth}
\includegraphics[scale=0.25]{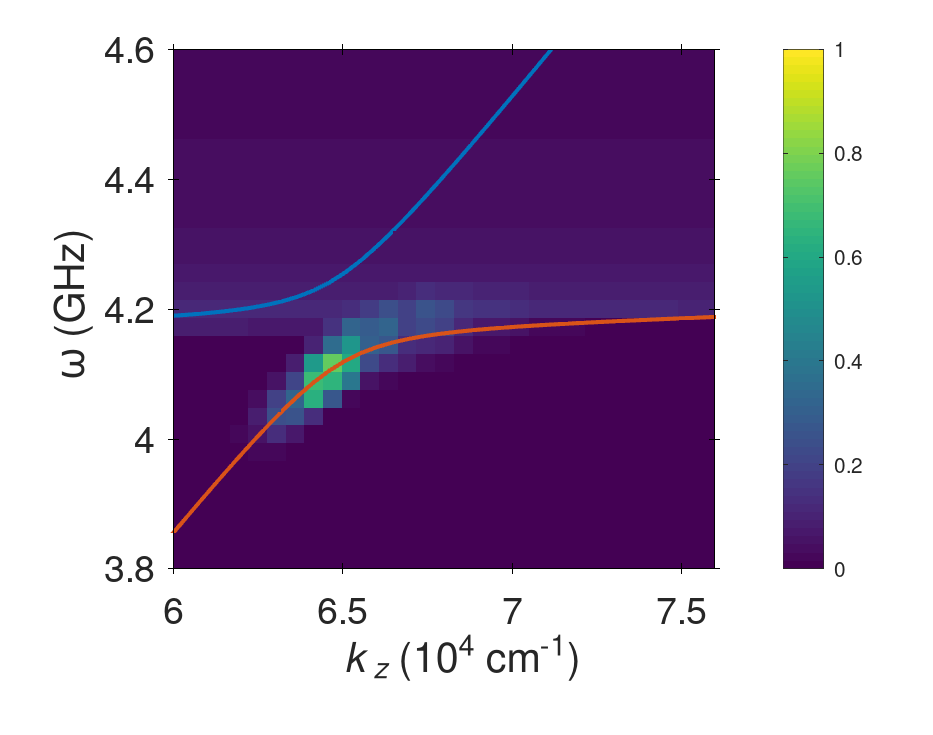}
\end{minipage}
\caption{
Magnon density $\rho_{ \kv m }$
as function of the wave vector $ \kv = \bm{e}_z k_z $ parallel to the external field
and the excitation frequency $\omega$,
normalized to the value of the magnon condensate.
The dispersion relations of the upper ($+$) and lower ($-$) magnetoelastic modes for
wave vectors parallel to the external magnetic field are indicated as blue and red lines respectively.  
The left peak in (a) is due to the magnon condensate.
(b) shows a close-up of the hybridization area.
}
\label{fig5}
\end{figure}

Our numerical results for the self-consistent steady state are shown in Fig.~\ref{fig5},
where the total magnon density 
\begin{equation}
\rho_{ \kv m } 
= \langle b^\dagger_\kv b_\kv \rangle
= \sum_{ r = \pm }
\left| \phi^1_{ \kv r } \right|^2 \left( n_{ \kv r } + \left| \Psi_{ \kv r } \right|^2 \right)
\end{equation}
is plotted as function of the wave vector $ \kv = \bm{e}_z k_z $ parallel to the external field
and the excitation frequency $\omega$. 
%
Apart from the condensate peak at the bottom of the magnon spectrum,
one clearly sees the emergence of a second sharp peak in the lower magnetoelastic mode
which  is located slightly below the bottom of the magnon spectrum in the hybridization area.
Despite the narrowness of this peak,
our simulations furthermore reveal that it is completely incoherent;
i.e., it is not associated with a finite condensate amplitude $\Psi_{ \kv - }$,
but only with the incoherent distribution $n_{ \kv - }$ of the magnetoelastic bosons. 
This peak  arises due to a bottleneck effect in the intermodal scattering across the hybridization gap, as discussed by Bozhko {\it{et al.}} \cite{Bozhko17}.

\begin{figure}[tb]
\centering
\includegraphics[scale=0.25]{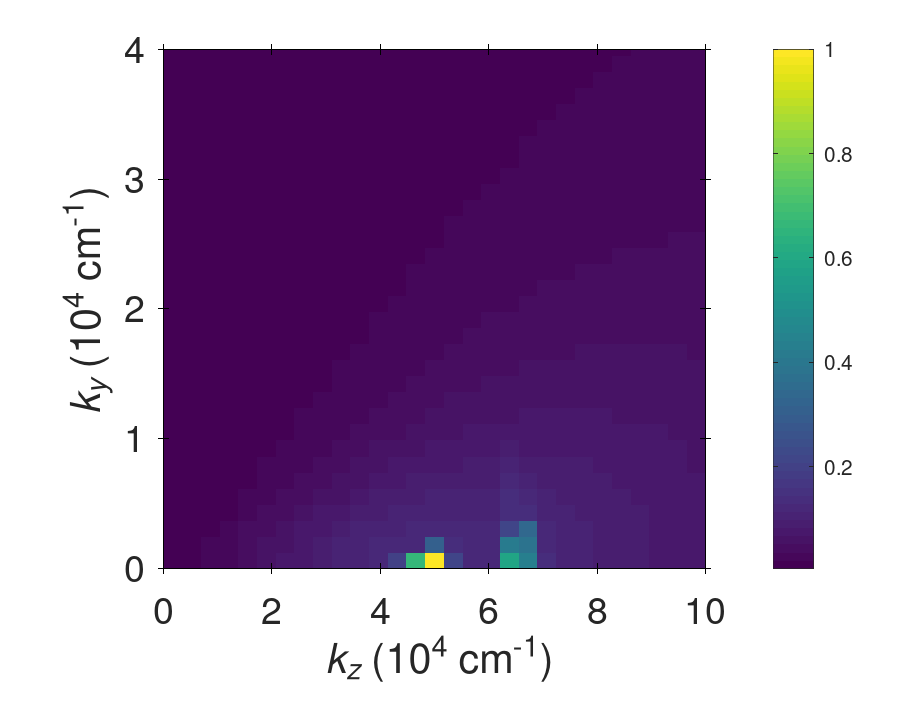}
\caption{
Magnon density $\rho_{ \kv m }$
in momentum space,
normalized as in Fig.~\ref{fig5}.
}
\label{fig6}
\end{figure}

The change in the magnon density in momentum space is displayed in Fig.~\ref{fig6},
which demonstrates that there is no significant deviation from the quasi-equilibrium state
away from the bottom of the magnon spectrum for wave vectors parallel to the external magnetic field.
In particular,
this means that the hybridization of magnons and phonons,
which is a continuous function of the angle $\theta_\kv$ between the wave vector and the external magnetic field,
is on its own not sufficient to observe an accumulation of magnetoelastic bosons.
Instead,
the near-degeneracy of this hybridization with the bottom of the magnon spectrum,
where the magnon condensate is located, is also necessary. 
Let us also point out that
the temperature and chemical potential of the thermal magnon cloud in this steady state are given by
$ T_m = 289.6 \, {\rm K} $ and $  \mu_m = 0.978 \, \epsilon_{ \kv_{\rm min} } $ respectively,
which is very close to the initial values.
Therefore the magnon distribution is virtually unaffected by the hybridization,
indicating the adequacy of our quasi-equilibrium ansatz \eqref{eq:superposition} 
for the incoherent distribution functions away from the hybridization area.

\begin{figure}[tb]
\centering
\begin{minipage}{16pt}
(a)
\end{minipage}
\begin{minipage}{0.91\linewidth}
\includegraphics[scale=0.25]{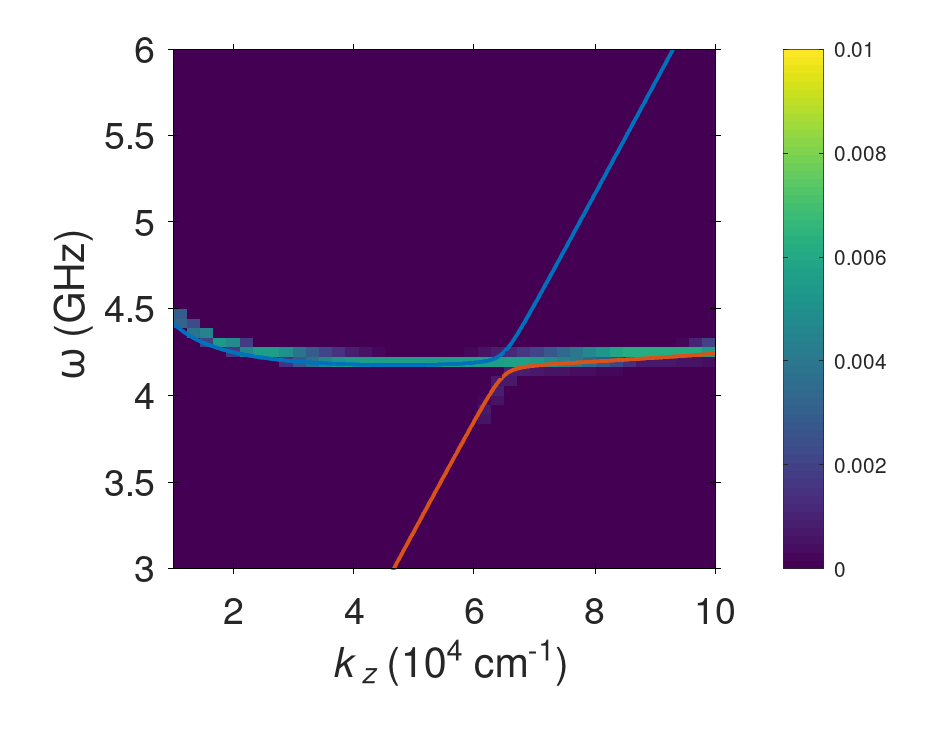}
\end{minipage}
\begin{minipage}{16pt}
(b)
\end{minipage}
\begin{minipage}{0.91\linewidth}
\includegraphics[scale=0.25]{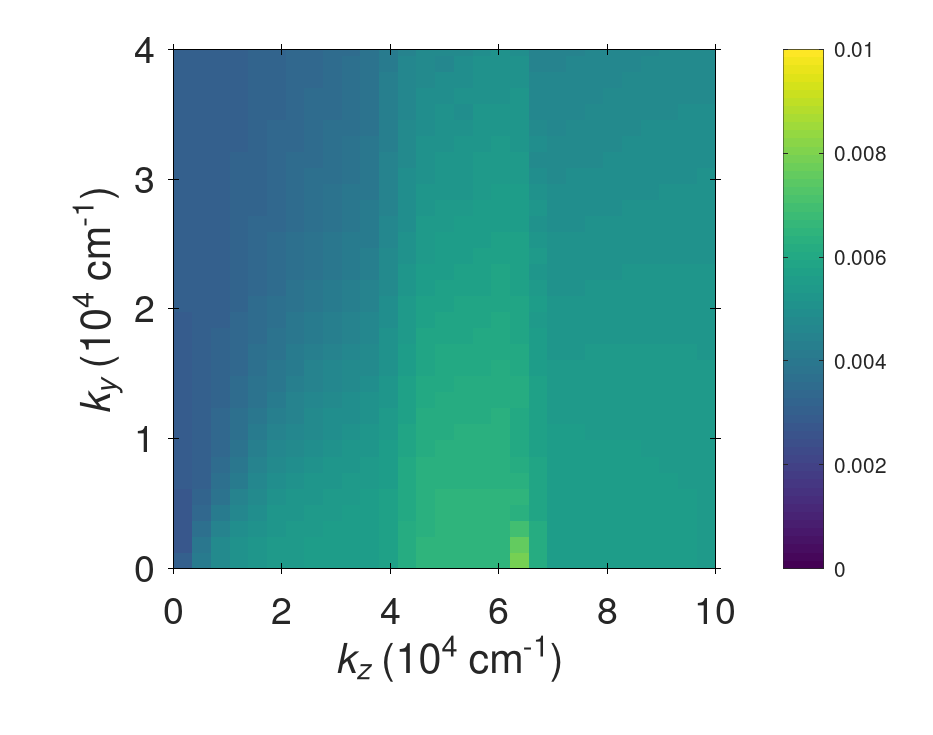}
\end{minipage}
\caption{
agnon density $\rho_{ \kv m }$
for a magnon gas that is not sufficiently pumped to establish a magnon condensate,
on the same scale as Fig.~\ref{fig5}.
(a) Magnon density 
as function of the wave vector $ \kv = \bm{e}_z k_z $ parallel to the external field
and the excitation frequency $\omega$.
The dispersion relations of the upper ($+$) and lower ($-$) magnetoelastic modes for
wave vectors parallel to the external magnetic field are indicated as blue and red lines, respectively.
(b) Magnon density in momentum space.
}
\label{fig7}
\end{figure}
To investigate the importance of the magnon condensate for the magnetoelastic accumulation,
we also show in Fig.~\ref{fig7} numerical results 
for the case that the magnon gas is not driven sufficiently strong to form a magnon condensate,
with  $  \mu_m = 0.75 \, \epsilon_{ \kv_{\rm min} } $.
Even in this case,
we observe a small bottleneck accumulation in the lower magnetoelastic mode, 
barely visible in Fig.~\ref{fig7}(b).
This is in agreement with Ref.~[\onlinecite{Bozhko17}],
where a magnetoelastic accumulation below the threshold of magnon condensation was reported.
However,
note the difference in scale:
While the magnetoelastic peak in Fig.~\ref{fig5} 
is of the same order of magnitude as the magnon condensate
and hence macroscopic,
it is only slightly enhanced compared to the thermal magnon gas
without a magnon condensate.
Thus, 
we conclude that the scattering of incoherent magnetoelastic bosons
with the nearly degenerate condensate amplitude 
is an important ingredient for the formation of a macroscopic magnetoelastic peak.


\subsection{Comparison with experiment}
\label{sec:experiment}

To further test the predictions of our simulations against experimental observations,
we have performed time- and wave vector-resolved Brillouin light scattering (BLS) spectroscopy \cite{Sandweg10} measurements of the  magnetoelastic accumulation at room temperature
in a $d = 6.7 \, \mu {\rm m}$ thick YIG film with dielectric coating.
An external magnetic field $\bm{H} = H {\bm{e}}_z$ of 145~mT is applied in-plane parallel to the $z$-axis.
Magnons are excited via a parallel parametric pumping \cite{Gurevich96,Serga12} pulse
of length $1500 \, {\rm ns}$.
During this process photons of the applied microwave field with frequency $f_p = 14 \, {\rm GHz}$ 
are splitting into two magnons with frequency $f_p/2$ and opposite wave vectors.
After the pumping pulse is switched off, the magnon gas 
rapidly thermalizes via number-conserving magnon-magnon scattering processes,
generating a finite chemical potential 
and eventually a magnon condensate at the bottom of the spectrum.

\begin{figure}[tb]
\centering
\begin{minipage}{9pt}
(a)
\end{minipage}
\begin{minipage}{0.95\linewidth}
\includegraphics[scale=0.25]{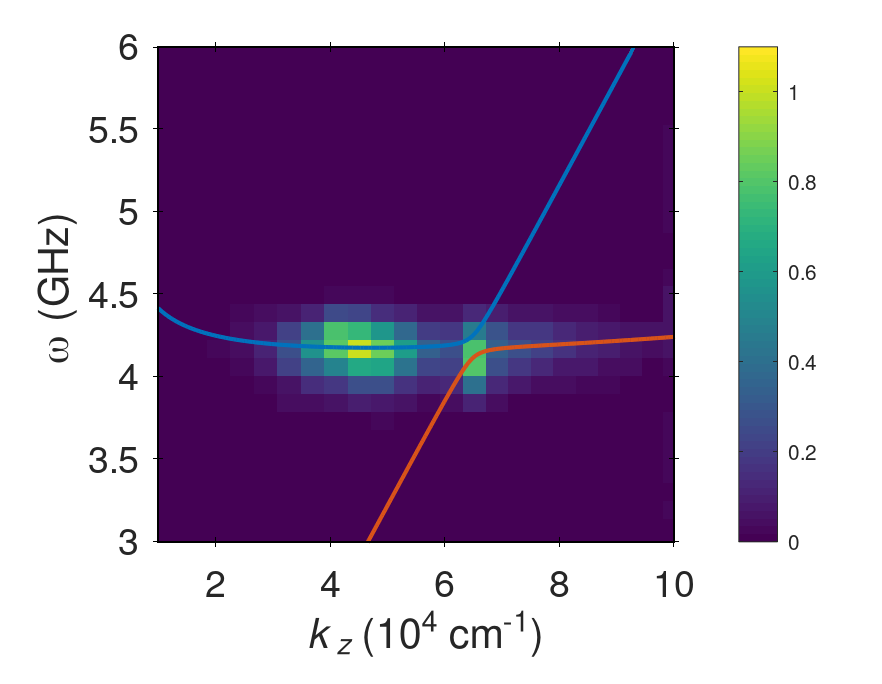}
\end{minipage}
\begin{minipage}{9pt}
(b)
\end{minipage}
\begin{minipage}{0.95\linewidth}
\includegraphics[scale=0.25]{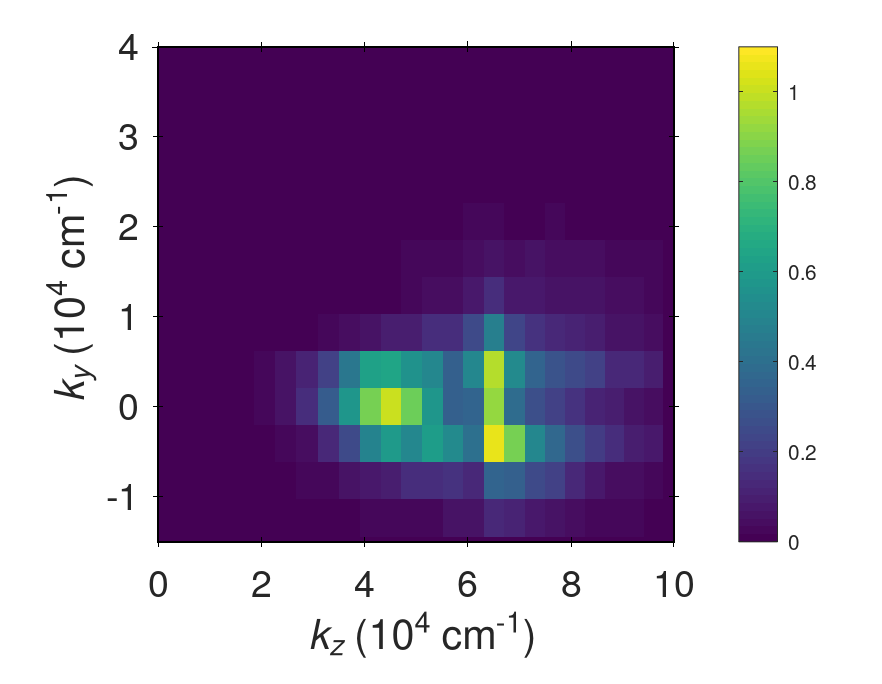}
\end{minipage}
\caption{
(a) Experimental magnon-phonon spectra and the population of the lowest magnon mode,
with the dispersion relations of the upper ($+$) and lower ($-$) magnetoelastic modes for
wave vectors parallel to the external magnetic field indicated as blue and red lines, respectively. 
(b) Wave vector-resolved magnon population.
Both plots are normalized to the value of the magnon condensate
to facilitate direct comparison with the theoretical prediction in Fig.~\ref{fig5}.
}
\label{fig8}
\end{figure}

Regarding the BLS spectroscopy experiment,
a probing laser beam is focused onto the YIG film 
and the frequency shift of the scattered light is analyzed with a tandem Fabry-P\'{e}rot interferometer. 
This method is selective for magnons with a certain wave vector 
depending on the incident angle of the probing laser.
Since the BLS setup is only sensitive to modes with a uniform profile along the film normal \cite{Bozhko20b},
we are only able to detect the magnon intensity in the lowest mode. 
The BLS intensity depending on the magnon wave vector and energy is shown in Fig.~\ref{fig8}.
%
Note that these measurements are in good agreement 
with the numerical results obtained from the solution
of the kinetic equations 
shown in Figs.~\ref{fig5} and \ref{fig6}. 
In particular,
the position of the peak in the magnetoelastic mode agrees very well with our theoretical predictions,
while its magnitude is of the same order as the magnon condensate.
The overall broader shape of the experimental distributions 
can at least partially be attributed to a lower resolution than in the numerical simulation.
As all qualitative features of the experiment are furthermore reproduced by our 
calculations,
we conclude that our non-equilibrium steady state solution of the kinetic equations
correctly describes the  relevant physics 
of the observed magnetoelastic accumulation in YIG.

\section{Summary and conclusions}

\label{sec:conclusion}

In this work we have studied the accumulation of magnetoelastic bosons 
-- hybrid quasiparticles formed by the coupling of  magnons and phonons --
in  an overpopulated magnon gas in YIG.
Starting from an effective spin Hamiltonian
and a phenomenological expression for the magnetoelastic energy,
we have derived quantum kinetic equations 
describing the dominant scattering mechanisms for the magnetoelastic bosons,
both for the incoherent quasiparticle distribution functions
and  for the condensate amplitudes.
Guided by the observation that the bulk of the magnon and phonon clouds
efficiently thermalize to their respective (quasi-)equilibria,
we have developed an efficient numerical  strategy which has enabled  us 
to self-consistently determine
the non-equilibrium steady state from the explicit solution of our  kinetic equations
without further approximation.
This self-consistent steady state solution has allowed us 
to reproduce the spontaneous accumulation of magnetoelastic bosons 
in a microscopic calculation.
For the first time,
we also presented 
a two-dimensional wave-vector resolved measurement of this accumulation in YIG,
which agrees well with our theoretical predictions.
In particular, our microscopic theoretical approach based on the
self-consistent solution of a quantum kinetic equation 
quantitatively  describes
the accumulation of quasiparticles
in the hybridization area of the lower magnetoelastic branch,
slightly below the bottom of the magnon spectrum.

Our study furthermore clarifies the importance of the magnon condensate 
for the accumulation of the magnetoelastic bosons: 
it turns out that
the existence of a magnon condensate strongly enhances the accumulation 
of magnetoelastic bosons.
Importantly,
we have also shown that despite the spectral narrowness of  the accumulation,
it  resides solely in the incoherent part of the distribution function 
and is thus not associated with a coherent state.
We expect that these findings will be helpful for future studies
of this intriguing phenomenon.

\section*{Acknowledgments}

This work was completed during a sabbatical stay of P.K. at the
Department of Physics and Astronomy at the University of California, Irvine.
P.K. would like to thank Sasha Chernyshev
for his hospitality.
A.R. acknowledges financial support by the Deutsche
Forschungsgemeinschaft (DFG, German Research Foundation) through Project No. KO/1442/10-1.
Partial support has been provided by the European Research Council within the Advanced Grant 694709 ‘‘SuperMagnonics: Supercurrents of Magnon Condensates for Advanced Magnonics’’ as well as financial support of the Deutsche Forschungsgemeinschaft (DFG, German Research Foundation) through the Collaborative Research Center ‘‘Spin+X: Spin in its collective environment’’ TRR – 173/2 – 268565370 (Project B04).

\appendix

\setcounter{equation}{0}

\renewcommand{\theequation}{A\arabic{equation}}

\renewcommand{\thesection}{\Alph{section}}

\section{EFFECTIVE MAGNON HAMILTONIAN FOR YIG} 

\label{app:magnon}

To make this work self-contained, we briefly review in this appendix
the derivation of the interaction Hamiltonian (\ref{eq:Hm4})  describing
two-body interactions between magnons in YIG. For a more detailed derivation see, for example,
Refs.~[\onlinecite{Kreisel09,Hick10,Hahn21}].
With the help of the Holstein-Primakoff transformation~\cite{Holstein40} the
effective spin-Hamiltonian \eqref{eq:H_eff}
can be expressed 
in terms of canonical boson operators $c_i$ and $c_i^{\dagger}$ as usual.
Expanding the resulting effective boson Hamiltonian in powers of  $1/S$ we obtain
\begin{equation}
 \mathcal{H}_m =\mathcal{H}_m^{(0)}+\mathcal{H}_m^{(2)}+\mathcal{H}_m^{(3)}+\mathcal{H}_m^{(4)}+\mathcal{O}(S^{-1/2}),
 \label{eq:H_1/S}
\end{equation}
so that $\mathcal{H}_m^{(n)} = {\cal O}( S^{ 2 - n / 2 } )$ contains the terms of order $n$ in the  
$c_i$ and $c_i^\dagger$.
Transforming to  momentum space,
\begin{equation}
c_i = \frac{1}{\sqrt{N}} \sum\limits_\kv e^{i\kv\cdot\bm{r}_i} c_\kv,
\end{equation}
where $N$ denotes the number of lattice sites in the $yz$-plane, 
we find that the quadratic part $\mathcal{H}_m^{(2)}$ of the Hamiltonian can be written as \cite{Hick10}
\begin{equation}
\mathcal{H}_m^{(2)} = \sum\limits_\kv\left[A_\kv c^\dagger_\kv c_\kv + \frac{B_\kv}{2}\left(c^\dagger_\kv c^\dagger_{-\kv} + c_{-\kv} c_\kv\right)\right],
\end{equation}
where
\begin{eqnarray}
A_\kv &=& h + S \left(J_{\bm{0}} - J_\kv\right) + S \left[D^{zz}_{\bm{0}} - \frac{1}{2}\left(D^{xx}_\kv + D^{yy}_\kv\right)\right],
\label{eq:Akdef} \nonumber\\\\
B_\kv &=& -\frac{S}{2} \left[D^{xx}_\kv - 2 i D^{xy}_\kv - D^{yy}_\kv\right],
 \label{eq:Bkdef}
\end{eqnarray}
and the Fourier transforms of the exchange and dipolar couplings 
are defined by
\begin{eqnarray}
J_\kv &=& \sum\limits_i \text{e}^{-i \kv \cdot \bm{r}_{i j}} J_{i j}, \\
D^{\alpha \beta}_\kv &=& \sum\limits_i \text{e}^{-i \kv \cdot \bm{r}_{i j}} D^{\alpha \beta}_{i j}.
\end{eqnarray}
As explained in Sec.~\ref{sec:magnons}, 
the cubic part ${\cal{H}}_n^{(3)}$
of the Hamiltonian can be neglected 
for our purpose because energy and momentum conservation cannot be fulfilled by the cubic interactions in the parameter regime of interest to us. 
Therefore we need only  the quartic part of the Hamiltonian, 
which reads~\cite{Hick10}
\begin{eqnarray}
\mathcal{H}_m^{(4)} &=& \frac{1}{N} \sum\limits_{\kv_1 \dots \kv_4} \delta_{\kv_1+\kv_2+\kv_3+\kv_4, 0} 
\nonumber\\*
&&
\times \Biggl[\frac{1}{\left(2!\right)^2} \Gamma^{\bar{c}\bar{c}cc}_{1, 2; 3, 4} c^\dagger_{-1} c^\dagger_{-2} c_3 c_4 \nonumber\\*
\nonumber\\*
&& \phantom{ \times } + \frac{1}{3!} \Gamma^{\bar{c}ccc}_{1; 2, 3, 4} c^\dagger_{-1} c_2 c_3 c_4
\nonumber\\*
&& \phantom{ \times } + \frac{1}{3!} \Gamma^{\bar{c}\bar{c}\bar{c}c}_{1, 2, 3; 4} c^\dagger_{-1} c^\dagger_{-2} c^\dagger_{-3} c_4 \Biggr], 
\label{eq:H4_ft}
\end{eqnarray}
where we abbreviate the momenta $\kv_i$ by $i$. 
The vertices are given by
\begin{subequations}
\begin{eqnarray}
\Gamma^{\bar{c}\bar{c}cc}_{1, 2; 3, 4} &=& -\frac{1}{2}\Biggl[J_{\kv_1+\kv_3} + J_{\kv_2+\kv_3} + J_{\kv_1+\kv_4} + J_{\kv_2+\kv_4} \nonumber\\*
&& + D^{zz}_{\kv_1+\kv_3} + D^{zz}_{\kv_2+\kv_3} + D^{zz}_{\kv_1+\kv_4} + D^{zz}_{\kv_2+\kv_4} \nonumber\\*
&& - \sum\limits_{i=1}^4 \left(J_{\kv_i} - 2 D^{zz}_{\kv_i}\right) \Biggr], \\
\Gamma^{\bar{c}ccc}_{1; 2, 3, 4} &=& \frac{1}{4} \left[D^{xx}_{\kv_2} - 2 i D^{xy}_{\kv_2} - D^{yy}_{\kv_2} + D^{xx}_{\kv_3}-2 i D^{xy}_{\kv_3} - D^{yy}_{\kv_3} \right.\nonumber\\*
&&\left. + D^{xx}_{\kv_4} - 2 i D^{xy}_{\kv_4} - D^{yy}_{\kv_4}\right], \\*
\Gamma^{\bar{c}\bar{c}\bar{c}c}_{1, 2, 3; 4} &=& \left(\Gamma^{\bar{c}ccc}_{4; 1, 2, 3}\right)^* .
\end{eqnarray}
\end{subequations}
The quadratic part $\mathcal{H}_m^{(2)}$ of the Hamiltonian can be diagonalized by the Bogoliubov transformation
to new canonical Bose operators $b_{\kv}$ and $b_{\kv}^\dagger$,
\begin{equation} \label{eq:Bogoliubov}
 \left( \begin{array}{c} c_{\bm{k}} \\ c^{\dagger}_{-\kv} \end{array} \right)
 = \left( \begin{array}{cc} u_{\bm{k}} & - v_{\bm{k}} \\
 - v_{\bf{k}}^{\ast} & u_{\bm{k}} \end{array} \right)
\left( \begin{array}{c} b_{\kv} \\ b^{\dagger}_{-\kv} \end{array} \right),
\end{equation}
where the  Bogoliubov coefficients are
\begin{subequations}
\begin{eqnarray}
u_\kv &=& \sqrt{\frac{A_\kv + \varepsilon_\kv}{2 \varepsilon_\kv}}, \\
v_\kv &=& \frac{B_\kv}{|B_\kv|} \sqrt{\frac{A_\kv - \varepsilon_\kv}{2 \varepsilon_\kv}},
\end{eqnarray}
\end{subequations}
and the magnon dispersion $\epsilon_\kv$ is given by
\begin{equation}
\epsilon_\kv = \sqrt{A_\kv^2 - |B_\kv|^2} .
\end{equation}
In terms of the new Bose operators, the quadratic part of the Hamiltonian has the form
\begin{equation}
\mathcal{H}_m^{(2)} = \sum\limits_\kv\left[\epsilon_\kv b^\dagger_\kv b_\kv + \frac{A_\kv-\epsilon_\kv}{2\epsilon_\kv}\right] .
\label{eq:H2mc}
\end{equation}
By neglecting the constant term in Eq.~\eqref{eq:H2mc} above, we arrive at Eq.~\eqref{eq:H2m}.
Finally, applying the Bogoliubov transformation \eqref{eq:Bogoliubov} to the quartic Hamiltonian \eqref{eq:H4_ft} 
and dropping the terms that do not conserve the magnon number yields the interaction Hamiltonian \eqref{eq:Hm4}, 
with the quartic vertex explicitly given by
\begin{eqnarray}
 U_{1, 2; 3, 4} &=& \phantom{..} \Gamma^{\bar{c}\bar{c}cc}_{1, 2; 3, 4} u_1 u_2 u_3 u_4 + \Gamma^{\bar{c}\bar{c}cc}_{1, 3; 4, 2} u_1 u_4 v_3 v_2 \nonumber\\*
 && + \Gamma^{\bar{c}\bar{c}cc}_{1, 4; 3, 2} u_1 u_3 v_4 v_2 + \Gamma^{\bar{c}\bar{c}cc}_{2, 3; 4, 1} u_2 u_4 v_3 v_1 \nonumber\\*
 && + \Gamma^{\bar{c}\bar{c}cc}_{2, 4; 3, 1} u_2 u_3 v_4 v_1 + \Gamma^{\bar{c}\bar{c}cc}_{3, 4; 2, 1} v_1 v_2 v_3 v_4 \nonumber\\*
 && - \Gamma^{\bar{c}ccc}_{4; 3, 2, 1} u_3 v_2 v_1 v_4 - \Gamma^{\bar{c}ccc}_{3; 4, 2, 1} u_4 v_2 v_1 v_3 \nonumber\\*
 && - \Gamma^{\bar{c}ccc}_{2; 3, 4, 1} u_2 u_3 u_4 v_1 - \Gamma^{\bar{c}ccc}_{1; 3, 4, 2} u_1 u_3 u_4 v_2 \nonumber\\*
 && - \Gamma^{\bar{c}\bar{c}\bar{c}c}_{2, 3, 4; 1} u_2 v_3 v_4 v_1 - \Gamma^{\bar{c}\bar{c}\bar{c}c}_{1, 3, 4; 2} u_1 v_3 v_4 v_2 \nonumber\\*
 && - \Gamma^{\bar{c}\bar{c}\bar{c}c}_{1, 2, 4; 3} u_1 u_2 u_3 v_4 - \Gamma^{\bar{c}\bar{c}\bar{c}c}_{1, 2, 3; 4} u_1 u_2 u_4 v_3 . \nonumber\\*
 \label{eq:vertex4}
\end{eqnarray}

\section{QUANTIZATION OF THE MAGNETOELASTIC ENERGY}

\label{app:phonon}

\setcounter{equation}{0}

\renewcommand{\theequation}{B\arabic{equation}}

In order to quantize the magnetoelastic energy \eqref{eq:E_me},
we first note that the (linear) symmetric strain tensor $X^{ \alpha \beta } ( \bm{r} )$
can be expressed in terms of the phonon displacement field $ \bm{X} ( \bm{r} ) $ as \cite{Landau70}
\begin{equation}
X^{ \alpha \beta } ( \bm{r} ) = \frac{ 1 }{ 2 } \left[
\frac{ \partial X^\alpha ( \bm{r} ) }{ \partial r^\beta } +
\frac{ \partial X^\beta ( \bm{r} ) }{ \partial r^\alpha }
\right] .
\end{equation}
Following the standard approach of expanding the displacement field 
in terms of the phonon creation and annihilation operators $a_{ \kv \lambda}^\dagger$ and $a_{ \kv \lambda }$
then yields
\begin{equation}
\bm{X} ( \bm{r} )
\to \frac{ 1 }{ \sqrt{ N } } \sum_{ \kv \lambda } e^{ i \kv \cdot \bm{r} } 
\frac{ a_{ \kv \lambda } + a_{ - \kv \lambda }^\dagger }{ \sqrt{ 2 \rho n \omega_{ \kv \lambda } } } \bm{e}_{ \kv \lambda } \; .
\end{equation}
where $n = 1/ a^3$ is the number density of ions
and $\rho \approx 5.17 \; \mbox{g}/ \mbox{cm}^2$
is the mass density of YIG.
The phonon polarization vectors 
$\bm{e}_{ \kv \lambda } = \bm{e}_{ -\kv \lambda }^*$ satisfy 
the orthogonality and completeness relations 
$\bm{e}_{ \kv \lambda }^* \cdot \bm{e}_{ \kv \lambda' } = \delta_{ \lambda \lambda ' }$ and
$\sum_\lambda \bm{e}_{ \kv \lambda } \bm{e}_{ \kv \lambda }^\dagger = {\bf 1}$.
In the thin film geometry of Fig.~\ref{fig1},
a convenient choice for the three polarization vectors is \cite{Rueckriegel14}
\begin{subequations}
\begin{align}
\bm{e}_{ \kv \parallel } =&
i \kv / | \kv | = i \left( \bm{e}_z \cos \theta_\kv + \bm{e}_y \sin \theta_\kv \right) , 
\\
\bm{e}_{ \kv \bot 1} =&
i \left( \bm{e}_z \sin \theta_\kv + \bm{e}_y \cos \theta_\kv \right) , 
\\
\bm{e}_{ \kv \bot 2 } =&
\bm{e}_x .
\end{align}
\end{subequations}

To leading order in $1/S$,
the local magnetization is quantized by replacing
\begin{subequations}
\begin{align}
M^x ( \bm{r} ) \to& 
\frac{ \mu n }{ \sqrt{ N } } \sum_{ \kv } e^{ i \kv \cdot \bm{r} } 
\sqrt{ \frac{ S }{ 2 } } \left( c_\kv + c_{ -\kv }^\dagger \right) ,
\\
M^y ( \bm{r} ) \to& 
\frac{ \mu n }{ \sqrt{ N } } \sum_{ \kv } e^{ i \kv \cdot \bm{r} } 
\frac{ 1 }{ i } \sqrt{ \frac{ S }{ 2 } } \left( c_\kv - c_{ -\kv }^\dagger \right) ,
\\
M^z ( \bm{r} ) \to& 
\mu n S .
\end{align}
\end{subequations}
With this prescription the classical magnetoelastic energy
$E_{\rm me}$ defined in Eq.~(\ref{eq:E_me})
is replaced by the quantized magnon-phonon Hamiltonian
${\cal H}_{ mp }^{ ( 2 ) } + {\cal O} ( 1 / S)$ with 
\begin{equation}
\mathcal{H}_{mp}^{(2)} 
= 
\frac{1}{2} \sum_{ \kv \lambda} \gamma_{ \kv \lambda }
\left( 
a_{ -\kv \lambda } + a_{ \kv \lambda }^\dagger
\right) c_\kv
+ {\rm h.c.} \; \; .
\end{equation}
For the thin-film geometry shown in Fig.~\ref{fig1} 
the hybridization vertices are  given by \cite{Rueckriegel14} 
\begin{subequations} \label{eq:gamma}
\begin{align}
\gamma_{ \kv \parallel } =&
i \frac{ B_\bot }{ \sqrt{ S \rho n \omega_{ \kv \lambda } } } \frac{ 2 k_y k_z }{ | \kv | }
= i \frac{ B_\bot }{ \sqrt{ S \rho n \omega_{ \kv \lambda } } }  | \kv | \sin \left( 2 \theta_\kv \right) , 
\\
\gamma_{ \kv \bot 1 } =&
i \frac{ B_\bot }{ \sqrt{ S \rho n \omega_{ \kv \lambda } } } \frac{ k_y^2 - k_z^2 }{ | \kv | }
= - i \frac{ B_\bot }{ \sqrt{ S \rho n \omega_{ \kv \lambda } } }  | \kv | \cos \left( 2 \theta_\kv \right) , 
\\
\gamma_{ \kv \bot 2 } =&
- i \frac{ B_\bot }{ \sqrt{ S \rho n \omega_{ \kv \lambda } } } k_z
= - i \frac{ B_\bot }{ \sqrt{ S \rho n \omega_{ \kv \lambda } } }  | \kv | \cos \theta_\kv  . 
\end{align}
\end{subequations}
In the last step,
we apply the Bogoliubov transformation (\ref{eq:Bogoliubov}) to the magnon operators,
which yields the magnon-phonon hybridization Hamiltonian given in Eq.~\eqref{eq:H2_mp},
with the transformed hybridization vertices
\begin{equation} \label{eq:Gamma}
\Gamma_{ \kv \lambda } = u_\kv \gamma_{ \kv \lambda } - v_\kv^* \gamma_{ - \kv \lambda }^* .
\end{equation}

\section{COLLISION INTEGRALS}

\label{app:collision}

\setcounter{equation}{0}

\renewcommand{\theequation}{C\arabic{equation}}

In this appendix we outline the derivation of the collision integrals
$I_{\kv \pm}$  and $\tilde{I}_{\kv \pm}$ in 
Eqs.~\eqref{eq:coll_inta} and (\ref{eq:coll_intb}).  Therefore we use the method developed
in Ref.~[\onlinecite{Fricke97}] which produces a systematic expansion of 
the collision integrals in powers of connected equal-time
correlation functions, see also Ref.~[\onlinecite{Hahn21}] for a recent application 
of this method in the context of YIG. 

\begin{figure}[tb]
\includegraphics[width=\linewidth]{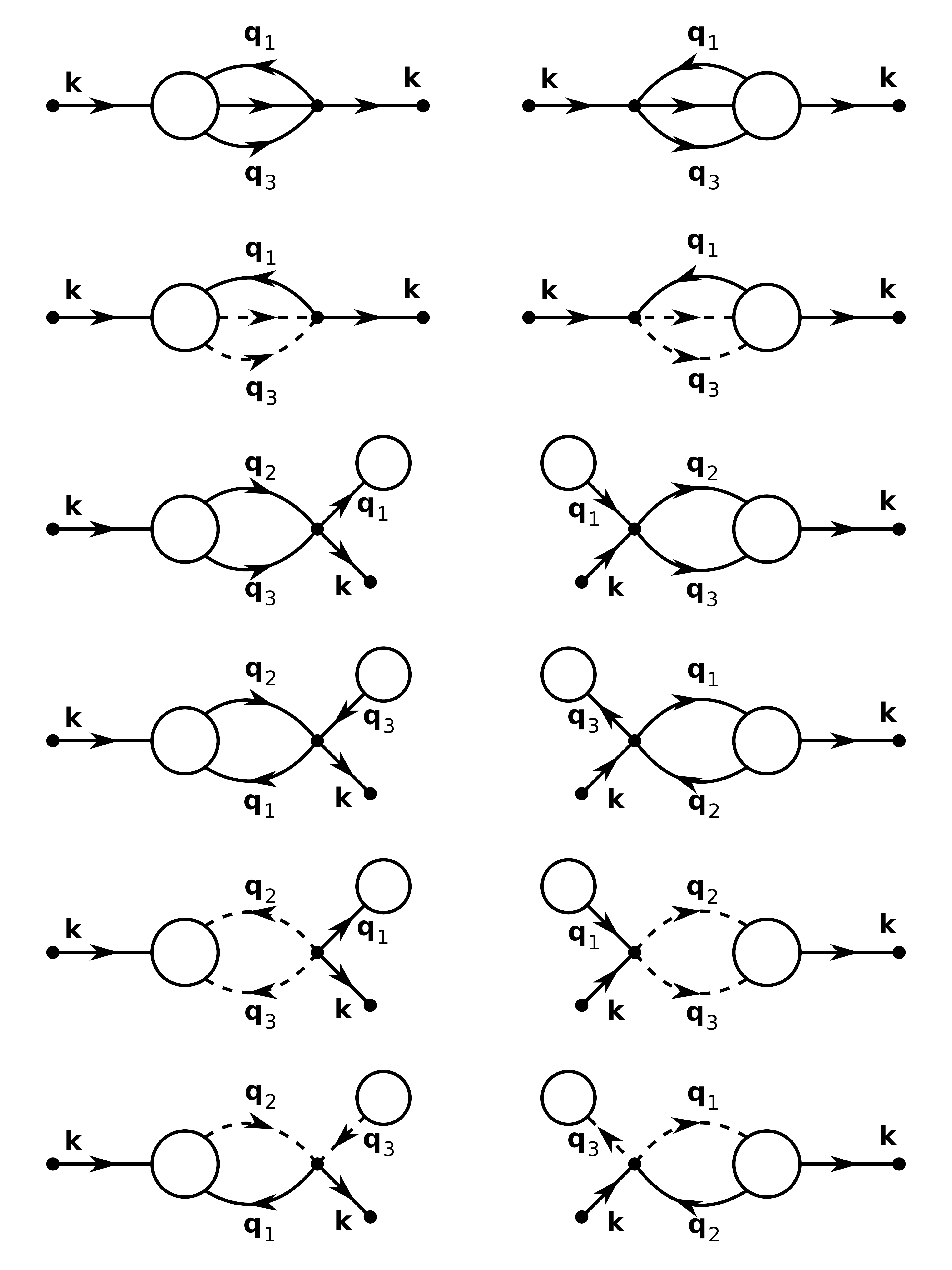}
\caption{
The diagrams contributing to the time evolution of the correlation 
$n_{ \kv + } = \langle \psi_{ \kv + }^\dagger \psi_{ \kv + } \rangle^c$
in an approximation where only the quartic vertices are retained.
Note that the diagrams used here 
differ from Feynman diagrams as they represent 
contributions to the differential equations for the correlations at a fixed time.
External vertices denote creation operators (outgoing arrows) or annihilation operators (incoming arrows) and internal vertices denote the bare interactions. Lines between the interaction vertices and external vertices represent connected correlations of order two where solid lines denote the upper ($+$) branch and dashed lines denote the lower ($-$) branch.
The circles represent connected correlations.
}
\label{fig9}
\end{figure}

Let us start with the collision integral $I_{\kv \pm}$ which controls the time-derivative
$\partial_t n_{\kv \pm}$  of the 
distribution  of the magnetoelastic modes.
A diagrammatic representation of the various  terms
contributing to this collision integral 
is shown in Fig.~\ref{fig9}.
Note that the circles in Fig.~\ref{fig9} represent the exact
equal-time correlations, while the black dots
represent the bare four-point vertices defined in
Eq.~(\ref{eq:U4def}).
\begin{figure}[tbp]
\centering
\includegraphics[width=0.8\linewidth]{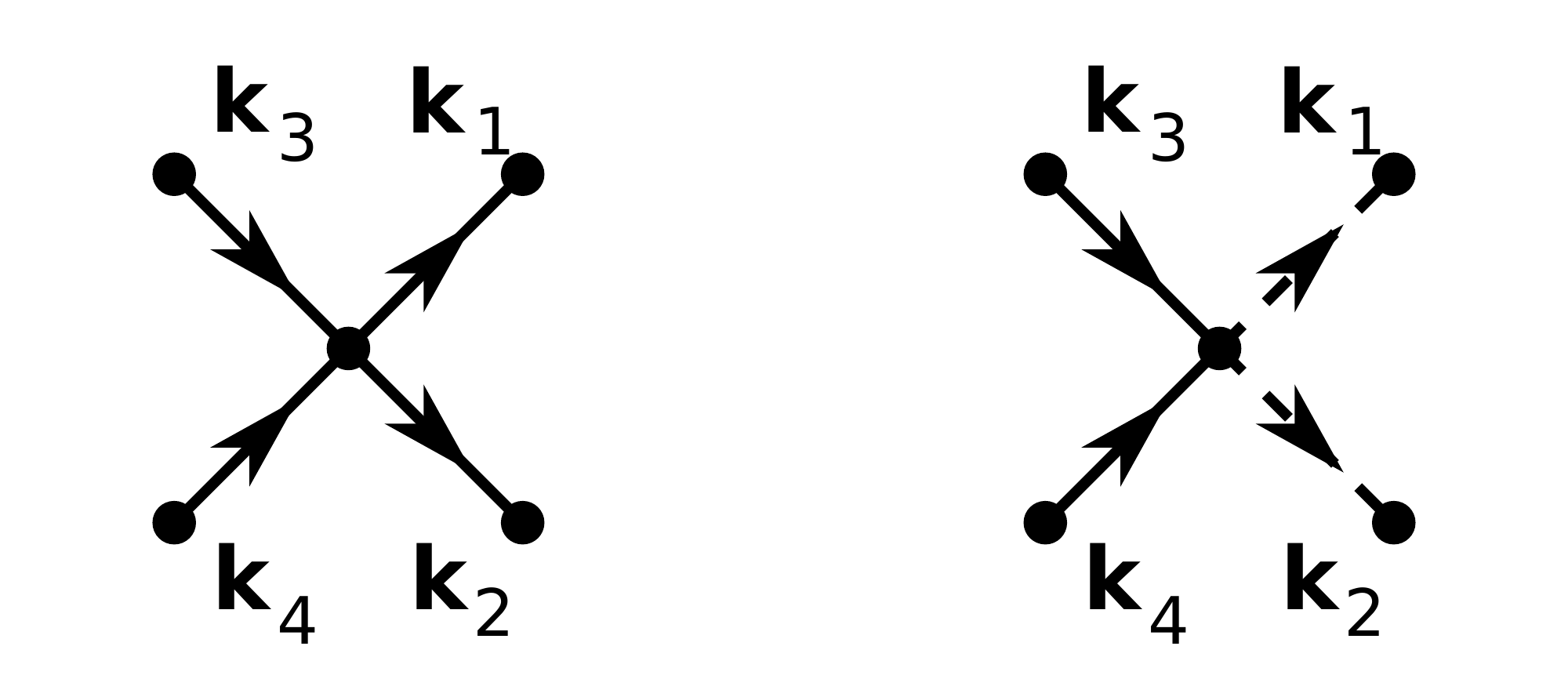}
\caption{The diagrams contributing to the time evolution of the correlations 
$\langle \psi_{ \kv_1 + }^\dagger \psi_{ \kv_2 + }^\dagger \psi_{ \kv_3 + } \psi_{ \kv_4 + } \rangle^c$ (left) and 
$\langle \psi_{ \kv_1 + }^\dagger \psi_{ \kv_2 + }^\dagger \psi_{ \kv_3 - } \psi_{ \kv_4 - } \rangle^c$ (right). The diagrams contain only the intramodal scattering vertex
$U_{1, 2; 3, 4}^{ + + + + } $ and the intermodal scattering vertex
$U_{1, 2; 3, 4}^{ + + - - } $.
Diagrams containing correlations are of higher order in the interaction vertices and are neglected here.
}
\label{fig10}
\end{figure}
\begin{figure}[tb]
\includegraphics[width=\linewidth]{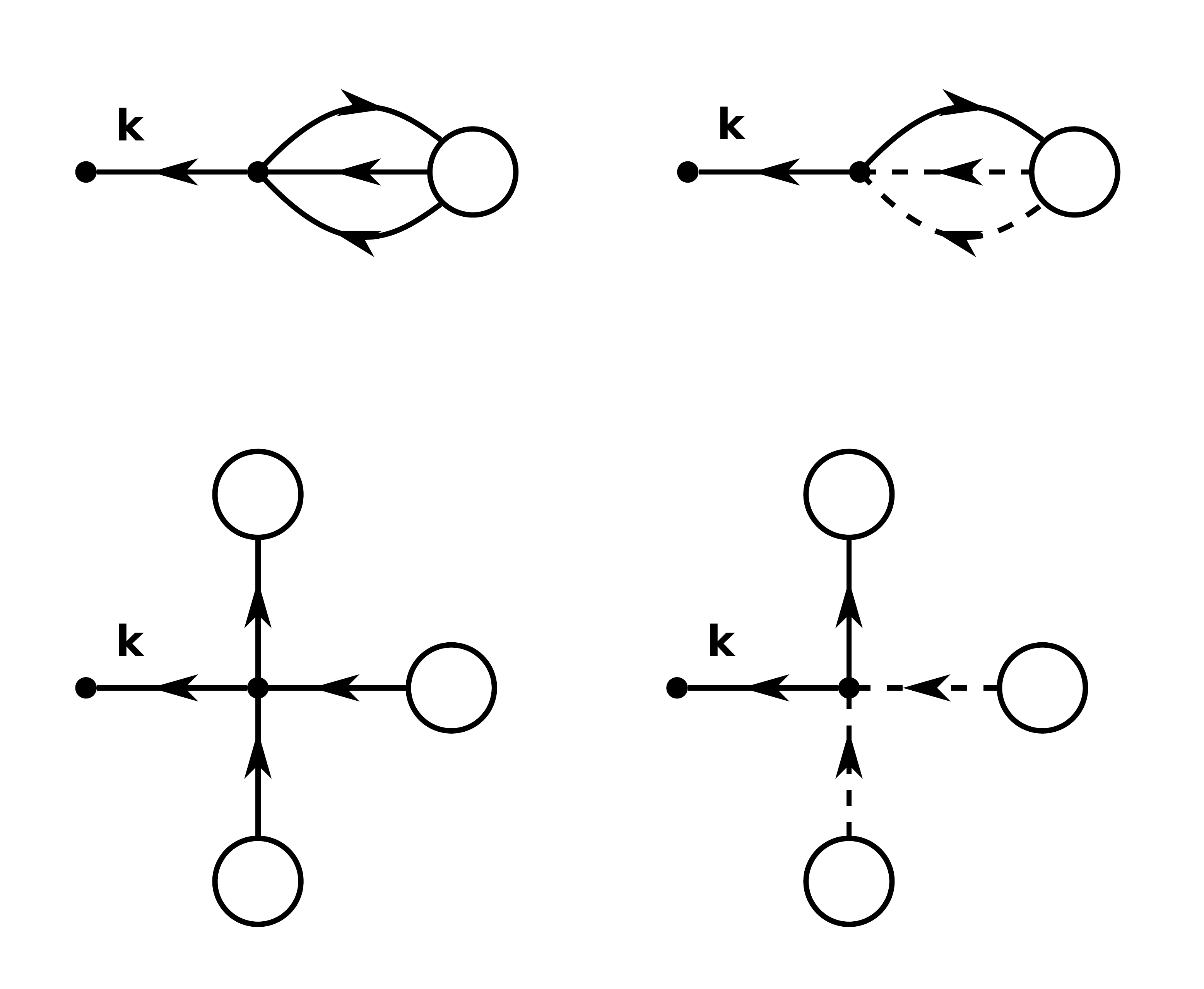}
\caption{
The diagrams contributing to the time evolution of the condensate amplitude 
$\Psi_{ \kv + } = \langle \psi_{ \kv + } \rangle$.
The graphical elements are defined in the caption of Fig.~\ref{fig9}.
The two diagrams in the second line represent the
Gross-Pitaevskii term on the left-hand side of the equation of motion
\eqref{eq:dynamics_Psi} for the
condensate density $\Psi_{\bm{k} \pm}$.
}
\label{fig11}
\end{figure}
\begin{figure}[tbp]
\includegraphics[width=\linewidth]{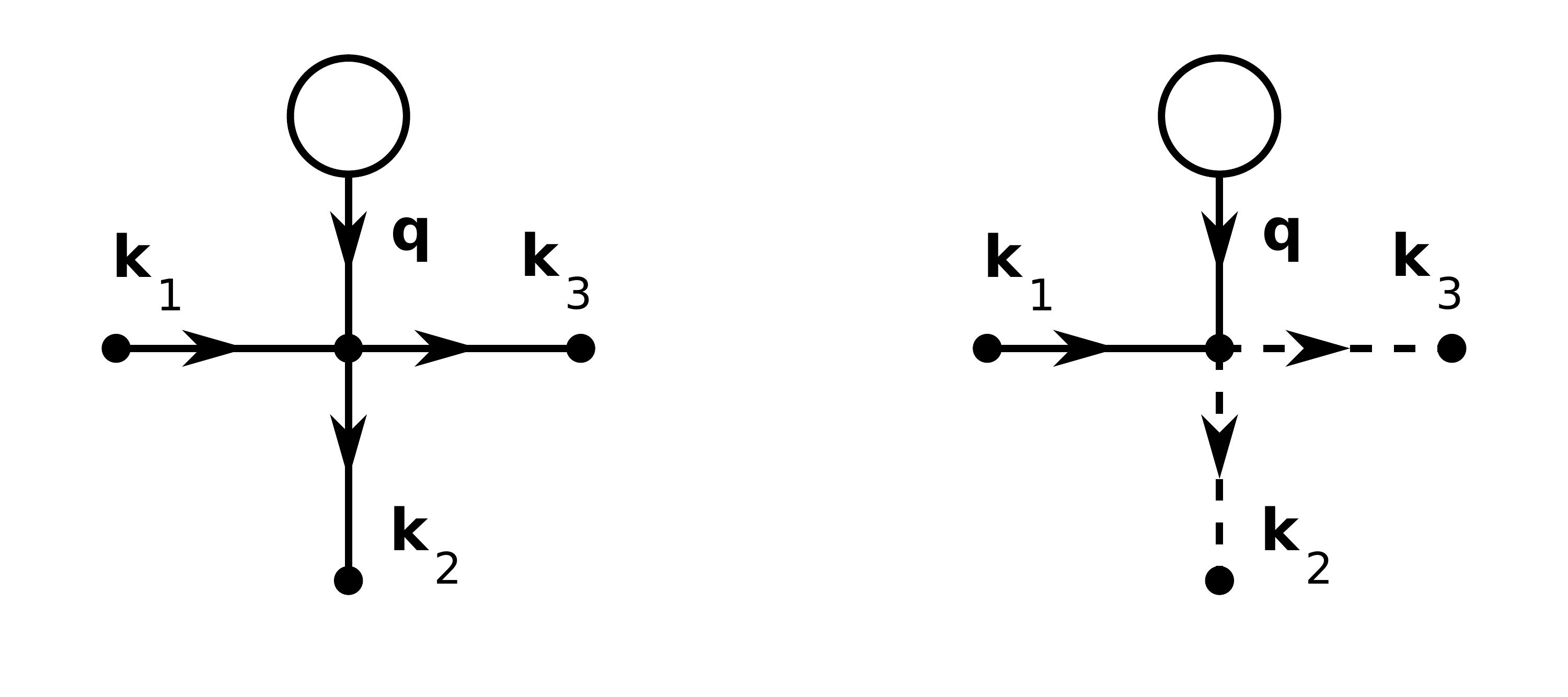}
\caption{
The diagrams contributing to the time evolution of the three-field correlations 
$\langle \psi_{ \kv_1 + } \psi_{ \kv_2 + }^\dagger \psi_{ \kv_3 + }^\dagger \rangle^c$ (left) and 
$\langle \psi_{ \kv_1 + } \psi_{ \kv_2 - }^\dagger \psi_{ \kv_3 - }^\dagger \rangle^c$ (right) contain the condensate amplitude and the intramodal scattering vertex $U_{1, 2; 3, 4}^{ + + + + } $ (left) and the intermodal scattering vertex $U_{1, 2; 3, 4}^{ + + - - } $ (right).
The symbols have the same meaning as in Fig.~\ref{fig9}.
Note that diagrams containing higher order correlations are neglected here because they lead to terms of higher order in the interaction vertices.
}
\label{fig12}
\end{figure}
These diagrams represent the following mathematical expression,
\begin{widetext}
\begin{eqnarray}
I_{ \kv + } 
&=& 
\frac{ i }{ 2N } \sum\limits_{\qv_1,\qv_2,\qv_3}
\delta_{\kv+\qv_1-\qv_2-\qv_3,0}
\left[  
U^{ + + + + }_{-\kv,-\qv_1;\qv_2,\qv_3} 
\langle
\psi_{ \qv_2 + }^\dagger \psi_{ \qv_3 + }^\dagger
\psi_{ \kv + } \psi_{\qv_1 + } 
\rangle^c 
+ 
U^{ + + - - }_{-\kv,-\qv_1;\qv_2,\qv_3} 
\langle
\psi_{ \qv_2 - }^\dagger \psi_{ \qv_3 - }^\dagger
\psi_{ \kv + } \psi_{\qv_1 + } 
\rangle^c 
-\mathrm{c.c.}
\right].\nonumber\\*
&&+ \frac{i}{2N} \sum\limits_{\qv_1,\qv_2,\qv_3}
\delta_{\kv+\qv_1-\qv_2-\qv_3,0}
\Biggl[
U^{ + + + + }_{-\kv,-\qv_1;\qv_2,\qv_3} 
\langle
\psi_{\qv_2 +}^\dagger \psi_{\qv_3 +}^\dagger
\psi_{\kv +}
\rangle^c
\langle
\psi_{\qv_1 +}
\rangle^c
+ 2
U^{ + + + + }_{-\kv,-\qv_1;\qv_2,\qv_3} 
\langle
\psi_{\qv_1 +}^\dagger
\psi_{\qv_2 +} \psi_{\kv +}
\rangle^c
\langle
\psi_{\qv_3 +}^\dagger
\rangle^c \nonumber\\*
&&+
U^{ + + - - }_{-\kv,-\qv_1;\qv_2,\qv_3} 
\langle
\psi_{\qv_2 -}^\dagger \psi_{\qv_3 -}^\dagger
\psi_{\kv +}
\rangle^c
\langle
\psi_{\qv_1 +}
\rangle^c
+ 2
U^{ + + - - }_{-\kv,-\qv_1;\qv_2,\qv_3} 
\langle
\psi_{\qv_1 +}^\dagger
\psi_{\qv_2 -} \psi_{\kv +}
\rangle^c
\langle
\psi_{\qv_3 -}^\dagger
\rangle^c
-\mathrm{c.c.}
\Biggr].
\label{eq:I_nU}
\end{eqnarray}
For the four-point and three-point correlations in this expression,
we use again  their equations of motion.
We will explicitly show only the calculations for the term shown in Fig.~\ref{fig9} contributing to the equation of motion of 
$\langle
\psi_{ \qv_2 + }^\dagger \psi_{ \qv_3 + }^\dagger
\psi_{ \kv + } \psi_{\qv_1 + } 
\rangle^c$ as an example which is,
\begin{eqnarray}
&&
\left[
\frac{d}{dt}
+i\left( 
E_{ \kv + }+E_{ \qv_1 + }-E_{ \qv_2 + }-E_{ \qv_3 + } 
\right)
\right] 
\langle
\psi_{ \qv_2 + }^\dagger \psi_{ \qv_3 + }^\dagger
\psi_{ \kv + } \psi_{\qv_1 + } 
\rangle^c 
\nonumber\\
&=& 
\frac{ i }{ 4N } 
U^{ + + + + }_{-\qv_2,-\qv_3;\kv,\qv_1}
\biggl(
\langle \psi_{ \kv + } \psi_{ \kv + }^\dagger \rangle^c
\langle \psi_{ \qv_1 + } \psi_{ \qv_1 + }^\dagger \rangle^c
\langle \psi_{ \qv_2 + }^\dagger \psi_{ \qv_2 + } \rangle^c
\langle \psi_{ \qv_3 + }^\dagger \psi_{ \qv_3 + } \rangle^c 
\nonumber\\
&& \hspace{80pt} -
\langle \psi_{ \kv + }^\dagger \psi_{ \kv + } \rangle^c
\langle \psi_{ \qv_1 + }^\dagger \psi_{ \qv_1 + } \rangle^c
\langle \psi_{ \qv_2 + } \psi_{ \qv_2 + }^\dagger \rangle^c
\langle \psi_{ \qv_3 + } \psi_{ \qv_3 + }^\dagger \rangle^c 
\biggr)
+ \dots
\nonumber\\
&=& 
\frac{ i }{ 4N } 
U^{ + + + + }_{-\qv_2,-\qv_3;\kv,\qv_1}
\left[ 
\left( 1 + n_{ \kv + } \right) \left( 1 + n_{ 1 + } \right) n_{ 2 + } n_{ 3 + }
- 
n_{ \kv + } n_{ 1 + } \left( 1 + n_{ 2 + } \right) \left( 1 + n_{ 3 + } \right)
\right] + \dots . 
\end{eqnarray}
The other contributions denoted by the dots contain three-point, four-point or six-point correlations,
which we neglect to leading order in the interaction.
As the contributions from the other diagrams have the same form the calculations are analogous for all terms.
We now integrate this equation to obtain the formal result
\begin{eqnarray}
\langle
\psi_{\qv_2 + }^\dagger
\psi_{\qv_3 + }^\dagger
\psi_{ \kv + } \psi_{ \qv_1 + }
\rangle^c 
&=&
\frac{ i }{ 4N } 
\int\limits_{t_0}^t dt'
\cos\left[
\left(
E_{ \kv + }+E_{ \qv_1 + }
-E_{ \qv_2 + }-E_{ \qv_3 + }
\right)
\left(t-t'\right)
\right] 
U^{ + + + + }_{ -\qv_2, -\qv_3;\kv,\qv_1}
\nonumber\\
&& \times
\left[ 
\left( 1 + n_{ \kv + } \right) 
\left( 1 + n_{ 1 + } \right) 
n_{ 2 + } n_{ 3 + }
- 
n_{ \kv + } n_{ 1 + } 
\left( 1 + n_{ 2 + } \right) 
\left( 1 + n_{ 3 + } \right)
\right]
+ \dots .
\label{eq:c4}
\end{eqnarray}
Inserting Eq.~\eqref{eq:c4} into Eq.~\eqref{eq:I_nU} then leads to
\begin{eqnarray}
I_{ \kv + } 
= \frac1{ 4N^2 } \sum\limits_{\qv_1,\qv_2,\qv_3} \delta_{\kv+\qv_1-\qv_2-\qv_3,0}
\left| U^{ + + + + }_{-\kv,-\qv_1;\qv_2,\qv_3} \right|^2
\int\limits_{t_0}^t dt' \cos \left[ 
\left( E_{ \kv + }+E_{ \qv_1 + }-E_{ \qv_2 + }-E_{ \qv_3 + } \right) \left(t-t'\right)
\right]
\nonumber\\
\times
\left[ 
\left( 1 + n_{ \kv + } \right) 
\left( 1 + n_{ 1 + } \right) 
n_{ 2 + } n_{ 3 + }
- 
n_{ \kv + } n_{ 1 + } 
\left( 1 + n_{ 2 + } \right) 
\left( 1 + n_{ 3 + } \right)
\right]
+ \dots .
\\[.25cm]
\underrightarrow{t_0 \rightarrow - \infty} \;
\frac{ \pi }{ 4N^2 } \sum\limits_{\qv_1,\qv_2,\qv_3} \delta_{\kv+\qv_1-\qv_2-\qv_3,0}
\left| U^{ + + + + }_{-\kv,-\qv_1;\qv_2,\qv_3} \right|^2
\delta \left( E_{ \kv + }+E_{ \qv_1 + }-E_{ \qv_2 + }-E_{ \qv_3 + } \right)
\nonumber\\
\times
\left[ 
\left( 1 + n_{ \kv + } \right) 
\left( 1 + n_{ 1 + } \right) 
n_{ 2 + } n_{ 3 + }
- 
n_{ \kv + } n_{ 1 + } 
\left( 1 + n_{ 2 + } \right) 
\left( 1 + n_{ 3 + } \right)
\right]
+ \dots ,
\end{eqnarray}
\end{widetext}
where in the last step we have taken the limit $t_0\rightarrow-\infty$.
In this way all terms entering the equation of motion for the one-particle distribution functions can be expressed in terms of the bare interaction vertices.

Finally, let us also give the diagrams contributing to the
collision integral $\tilde{I}_{\kv \pm}$ in Eq.~\eqref{eq:coll_intb}
which appears in the equation of motion \eqref{eq:dynamics_Psi} for
the condensate density $\Psi_{\bm{k} \pm}$.
The diagrams in the first line of Fig.~\ref{fig11}  
represent the contributions to  
the equation of motion for the condensate density
$\Psi_{\bm{k} \pm}$ involving  higher-order correlations.
On the other hand, the diagrams in the second line of Fig.~\ref{fig11} correspond to the Gross-Pitaevskii term which is not included in the collision 
integral in Eq.~\eqref{eq:dynamics_Psi}.
To lowest order in the interaction, the equation of motion for the  
three-point correlations 
in the diagrams of the first line of Fig.~\ref{fig12} 
can be expressed again in terms of the bare four-point vertices as shown
in Fig.~\ref{fig12}.


\begin{thebibliography}{99}
\bibitem{footnote_DE}
The negative slope of the magnon dispersion for wave vectors perpendicular
to the magnetic field for very long wavelengths
is an artifact of the thin film approximation.
For thicker films in the $\mu$m range, 
the thin film approximation fails to correctly account 
for the hybridization of different 
low-energy thickness modes for angles $ \theta_\kv \gtrsim 45^\circ $,
which results in the shallow minimum observed in Fig.~\ref{fig2}.
This inaccuracy of the thin film approximation is well-known
and has been discussed in detail in Ref.~[\onlinecite{Kreisel09}].
For our purposes,
it is of no consequence because we are ultimately only interested in the 
minimum of the magnon dispersion for $ \theta_\kv = 0^\circ $
and the magnon-phonon hybridization,
which occurs at slightly larger wave vectors.
\bibitem{footnote_muc}
Note that in our  non-equilibrium setup
the chemical potential $\mu_c$ 
of the magnons at the minimum of the dispersion (which form the condensate)
is in general different from the chemical potential $\mu_m$
of the other magnons. For our calculation we fix
$\mu_c$ and then determine  $\mu_m$ and the magnon temperature $T_m$ self-consistently.
\end{thebibliography}
\end{document}